\documentstyle[preprint,aps,prb]{revtex}

\begin{document}
 
\draft

\newcommand{\U}{{\cal U}} 
\def\G#1#2{{\cal G}^0_\Delta(#1,#2;\beta)} 
\def\GE#1#2{{{\cal G}^0_\Delta}_E(#1,#2;\beta)} 
\newcommand{\K}{{\cal K}^0_\Delta}
\newcommand{\g}{g^0_\Delta} 
\newcommand{\Ib}{\int_{-\beta/2}^{\beta/2}}
\def\intq#1{\int \frac{d^4#1}{(2\pi)^4}}
\def\intt#1{\int \frac{d^3#1}{(2\pi)^3}}
\def\intf#1{\int \frac{d#1_0}{2\pi}}
\newcommand{\bsig}{\mbox{\boldmath $\sigma$}}
\newcommand{\btau}{\mbox{\boldmath $\tau$}}
\newcommand{\bLam}{\mbox{\boldmath $\Lambda$}}
\newcommand{\brho}{\mbox{\boldmath $\rho$}}
\newcommand{\bgam}{\mbox{\boldmath $\gamma$}}
\newcommand{\bfeta}{\mbox{\boldmath $\bfeta$}}
\def\bfm#1{{\mbox{\boldmath $#1$}}}
\newcommand{\k}{{k_F}}
\newcommand{\llan}{\left<}
\newcommand{\rran}{\right>}
\renewcommand{\thefootnote}{\alph{footnote}}
\newcommand{\bhn}{{\bf\hat n}}
\newcommand{\beq}{\begin{equation}}
\newcommand{\eeq}{\end{equation}}
\newcommand{\bea}{\begin{eqnarray}}
\newcommand{\eea}{\end{eqnarray}}

\newcommand{\ket}[1]{| {#1} \rangle}
\newcommand{\bra}[1]{\langle {#1} |}
\newcommand{\ave}[1]{\langle {#1} \rangle}
\newcommand{\bce}{\begin{center}}
\newcommand{\ece}{\end{center}}
\newcommand{\ab}{{\alpha\beta}}
\newcommand{\cd}{{\gamma\delta}}
\newcommand{\dc}{{\delta\gamma}}
\newcommand{\ac}{{\alpha\gamma}}
\newcommand{\bd}{{\beta\delta}}
\newcommand{\abc}{{\alpha\beta\gamma}}
\newcommand{\abcd}{{\alpha\beta\gamma\delta}}
\newcommand{\bsigma}{\mbox{\boldmath $\sigma$}}
\newcommand{\bpipi}{\mbox{\boldmath $\pi\pi$}}
\newcommand{\bss}{\bsigma\!\cdot\!\bsigma}
\newcommand{\btt}{\btau\!\cdot\!\btau}
\newcommand{\bnab}{\mbox{\boldmath $\nabla$}}
\newcommand{\fpi}{f_\pi}
\newcommand{\mpi}{m_\pi}
\newcommand{\pit}{\textit{\texttt{p}}}
\newcommand{\qit}{\textit{\texttt{q}}}
\newcommand{\Tr}{{\mbox{\rm Tr}}}
\newcommand{\half}{{1\over 2}}
\newcommand{\quart}{{1\over 4}}
\newcommand{\cl}[1]{\begin{center} {#1} \end{center}}
\newcommand{\ba}{\begin{array}}
\newcommand{\ea}{\end{array}}
\newcommand{\bpi}{\mbox{\boldmath $\pi$}}
\newcommand{\bphi}{\mbox{\boldmath $\phi$}}
\newcommand{\bthet}{\mbox{\boldmath $\theta$}}
\newcommand{\bmu}{\mbox{\boldmath $\mu$}}
\newcommand{\qcd}{{\it QCD}}
\newcommand{\ie}{{\sl i.e.~}}
\newcommand{\etal}{{\sl et al.~}}
\newcommand{\etc}{{\sl etc.~}}
\newcommand{\rhs}{{\sl rhs~}}
\newcommand{\lhs}{{\sl lhs~}}
\newcommand{\eg}{{\sl e.g.~}}
\newcommand{\bfR}{\mbox{\boldmath ${\cal R}$}}
\newcommand{\bfL}{\mbox{\boldmath ${\cal L}$}}
\newcommand{\bfM}{\mbox{\boldmath ${\cal M}$}}
\def\slash#1{\rlap{\hskip1pt /}#1}

\bibliographystyle{unsrt}
\vskip -3cm 
\preprint{ECT*/06-98}
\vskip 2cm

\title
{\bf Similarity Renormalization, Hamiltonian Flow\\ 
\vskip-.5cm
Equations, and Dyson's Intermediate Representation}

\author{\vskip-1.4cm T.S. Walhout\footnote[2]{
e-mail address: walhout@ect.unitn.it}}
\address{\vskip-.1cm European Center for Theoretical
Studies in Nuclear Physics and Related Areas\\
 \vskip-.2cm Villa Tambosi, Strada delle Tabarelle 286;
\,  I-38050 Villazzano (Trento), Italy}

 
\maketitle
\tightenlines

\begin{abstract} 
A general framework is presented for the
renormalization of Hamiltonians via a similarity
transformation. Divergences in the similarity
flow equations may be handled with
dimensional regularization in this
approach, and the resulting
effective Hamiltonian is finite since states
well-separated in energy are uncoupled.
Specific schemes developed several years ago by 
G{\l}azek and Wilson and
contemporaneously by Wegner correspond to
particular choices within this framework, and
the relative merits of such choices are
discussed from this vantage point. 
It is shown that a scheme for the
transformation of Hamiltonians introduced by
Dyson in the early 1950's also corresponds to
a particular choice within the similarity
renormalization framework, and it is argued
that Dyson's scheme is preferable to the others
for ease of computation. As an example, it is
shown how a logarithmically confining potential
arises simply at second order in light-front
QCD within Dyson's scheme, a result found previously
for other similarity renormalization schemes.
Steps toward higher order and nonperturbative
calculations are outlined.  In particular, a 
set of equations analogous to Dyson-Schwinger
equations is developed.
\end{abstract}
 
\pacs{11.10.Ef;11.10.Gh;12.38.-t}


\renewcommand{\thefootnote}{\fnsymbol{footnote}}
 
\narrowtext
 
\section{Introduction}

While particle physics has been and remains to be guided
by the desire to understand phenomena at ever shorter
distances, one may nevertheless argue that the most
vital challenge for quantum field theory has little to do with the
physics at Planck length scales or a theory of everything (TOE).
This challenge is how to solve strongly interacting quantum
field theories, which inevitably amounts to understanding in
a nonperturbative way how
a theory designed to fit the physics of a smaller scale
will determine the phenomena observed at greater length
scales. The prototype is Quantum Chromodynamics (QCD), but the
problem has existed since soon after the development
of the renormalization program for Quantum Electrodynamics,
when physicists
began to study strongly interacting meson theories; and
the problem must presumably be confronted by anyone who wants
to show that any given TOE is indeed a TOE.

In ordinary quantum mechanics, strongly interacting
systems may be handled by a variety of techniques based
upon a Hamiltonian formulation of the theory in question.
Relativistic quantum field theory, on the other hand, is
studied almost exclusively in the Lagrangian formulation
nowadays. Lagrangian methods are favored because they
better allow one to keep symmetries of the theory
manifest. However, this generally entails a sacrifice
in calculational flexibility, with the result that one
is limited to applications for which perturbative or
semiclassical methods are valid (or at least
hoped to be valid). This paper takes the view that a Hamiltonian
formulation will provide a better starting point for the
study of strongly interacting quantum field theories.
It was this motivation, and the difficulties in its
realization, that eventually led Wilson to formulate his
version of the renormalization group.\cite{W65}

The renormalization group is most often
used in a very restricted way in quantum field
theory. Typically, renormalization is used as a
tool only to cover up infinities, to select
out well-behaved (``renormalizable'') theories, and
to make finite redefinitions of couplings so as to
reorganize perturbative expansions in a better way.
Renormalization transformations that produce more
complicated interaction terms are generally
avoided. This is because such transformations
usually break at least some symmetries and complicate
perturbative expansions. For a theory like QCD,
however, perturbation theory is not powerful enough to
calculate low-energy phenomena, for
the lowest orders of a perturbative expansion
look nothing like what finds from low-energy experiments. In such
a case, it might be worthwhile to consider
a renormalization group transformation that
brings new interactions, even if such a transformation
hides symmetries that in the usual perturbative
scheme remain manifest. The idea is that such a renormalization
group transformation can be designed so that the new interactions
it generates will contain the most important physics.

In fact, this is the major motivation behind the
similarity transformation of Ham\-il\-ton\-ians.\footnote[2]{Matrix
elements of the transformed Hamiltonian $H'$
are ``similar'' to the original Hamiltonian $H$ if
$H'=P H P^{-1}$. If $H'$ is to be hermitian, we
need $P^\dag = P^{-1}$, and so the similarity transformations
described here will all be unitary transformations.}
 The object of the similarity transformation is not primarily
to produce a renormalized Hamiltonian, but rather
to produce a Hamiltonian that can be dealt with in a way
such that even first approximations in its
analysis provide a qualitatively
correct description of the physics. That the similarity
transformation can be used to renormalize a Hamiltonian
that contains divergences is just a particular result of
this more general requirement. The practical
manifestation of this basic requirement is that the similarity
transformation produces a Hamiltonian where only states
of nearby energy contribute to phenomena at
a given energy scale. Thus G{\l}azek and Wilson\cite{GW} first
introduced their similarity transformation to deal with
divergences in light-front field theory, but also with
the idea of using it to help transform the QCD Hamiltonian
into a form for which a lowest-order approximation begins
to look like a constituent quark model.\cite{WWH} 
Wegner,\cite{We} 
on the other hand, independently, but at roughly the same time, 
introduced a set of Hamiltonian flow equations to deal with
many-body problems which have no need of the
renormalization of divergences.

We shall follow the presentation of the similarity renormalization
scheme given in Ref. [3], generalizing it in the
following. The transformations of Wegner and of G{\l}azek and
Wilson will be seen to follow from more specific choices within
our formalism. These three authors are generally given credit
for introducing the similarity transformation to field
theory. Interestingly enough, however, soon after the
development of renormalization theory, Dyson himself in
a difficult, fascinating, and seemingly universally
ignored series of papers\cite{D} developed what he called
the intermediate representation,
which results from a transformation
that matches all the requirements of the similarity
transformations developed by G{\l}azek, Wilson, and Wegner.
This is of more than just historical interest\cite{S}: we shall
see that Dyson's scheme has some features that lead it
to be preferred over the more recent schemes.

The outline of the paper is as follows. This section has been
an introduction. In the next section we provide a general
interaction representation formulation of the similarity
renormalization of Hamiltonians with respect to differences
in unperturbed energies. We show how the schemes of Wilson
and G{\l}azek and of Wegner correspond to specific choices
in our framework and argue that one might be able to make
a more manageable specific choice of transformation than
these. In Section III we discuss the perturbative solution
of the similarity flow equations for these schemes and conclude
that the corresponding diagrammatic expansion, which
corresponds to that of old-fashioned time-ordered perturbation
theory, leaves something to be desired in terms of
computational simplicity.

This leads to a detailed discussion of Dyson's intermediate
representation, which we show corresponds to a specific
choice of perturbative scheme in the similarity renormalization
framework. The beauty of Dyson's scheme is that it may be
represented diagrammatically in terms of generalized
Feynman graphs --- generalized in that amplitudes may be
off energy shell.  The difficulty comes in demonstrating
this, and we dedicate some space to summarizing Dyson's
presentation. Once the diagrammatic expansion is 
developed, we conclude that Dyson's scheme
is much simpler computationally --- at least for perturbative
solutions --- than those of Wilson, G{\l}azek and Wegner. We
end this long section with an example calculation taken
from QCD.

In Section IV we formulate the solution of the similarity
flow equations in terms of general $n$-particle ``dressed''
vertices. The resulting equations, which look much like
Dyson-Schwinger equations, are of second order in these
vertices; and we show how a specific choice of similarity
transformation --- a modification of Wegner's scheme ---
can guarantee that these dressed vertices are joined by
Feynman propagators, which therefore permits a Wick rotation
to Euclidean space. We show how analogous equations can also
be derived in Dyson's scheme and that these equations are
linear in the dressed vertices. Finally, we give a specific
example of how one may make approximations to sum a set of
diagrams similar to those summed in the Bethe-Salpeter
equation.  

We shall refer to a generic theory of fermionic and
vector bosonic fields throughout the paper. To provide
a flavor of the type of calculation involved here,
we give specific examples for QCD in light-front coordinates and
light-cone gauge, but we shall only
refer to other work that presents the Feynman rules for these
particular calculations. In Section III, 
we show that the similarity-transformed
Hamiltonian contains a potential that rises logarithmically
for large quark-antiquark separations, a result obtained
previously in other similarity renormalization schemes. In
Section IV we discuss a ladder-type summation of these
potentials. These examples are only meant to be suggestive.
More detailed applications of our similarity renormalization
framework will be undertaken in future papers.

Section V concludes the paper with a discussion
of the ideas presented here. This last section is not just a
summary but rather is intended to be an integral part of the
paper, for it provides further motivation that could not
properly be given before the scheme itself was developed.
In particular, the significance of the similarity renormalization
scheme for applications in
light-front field theory and the motivation for the use
of light-front coordinates in field theory in this light
are discussed at length.
 
\section{Similarity Renormalization Scheme}

\subsection{General formulation}

We begin with a regularized
field theoretical Hamiltonian $H$, derived
in the usual way from a renormalizable
relativistic Lagrangian. We write $H=H_0+H_{int}$, with
$H_0$ the usual free Hamiltonian. 
Let us assume that $H$ describes
processes which at high energies may be well approximated
by treating the interaction $H_{int}$ perturbatively,
but that its low-energy processes
and its bound states are essentially nonperturbative.
Clearly, the problem here is that $H_0$ is
a poor first approximation for calculating low-energy
phenomena.
The similarity renormalization scheme seeks to change
the original
Hamiltonian into a form that is more suitable for
calculations over a wider range of energies. This is done by
introducing an energy scale $\sigma$
via a similarity transformation $S_\sigma$
(which will always be unitary here). Roughly speaking,
the transformed Hamiltonian $H_\sigma = S_\sigma
H S_\sigma^\dagger$
will depend on $\sigma$, the similarity scale,
as follows: processes described by $H$ which transfer an energy
much greater than $\sigma$ will be absorbed into the structure of
$H_\sigma$, so that $H_\sigma$ will only describe processes
that correspond to energy transfers of order $\sigma$ or less.

Thus we call $H$ the undressed Hamiltonian and $H_\sigma$ the
dressed Hamiltonian.
The undressed Hamiltonian is renormalized; that is, the
undressed Hamiltonian is the sum of a bare regularized
Hamiltonian $H_B$ and a set of counterterms $H_C$ 
that removes any dependence in
physical quantities upon the regulators. The structure
of these counterterms must be determined as
one solves for the dressed Hamiltonian $H_\sigma$.
In particular, the counterterms in $H$ ensure that
$H_\sigma$ will remain finite as the regulators are 
removed. The dressed Hamiltonian $H_\sigma$
thus cannot produce infinities; and so the similarity
transformation $S_\sigma$ is finite, provided that
it avoids singularities corresponding to small
energy denominators. This latter restriction shall
be discussed presently.

To define such a transformation, we write
the unitary operator $S_\sigma$ in terms of the
antihermitian generator of infinitesmal similarity
scale transformations $T_\sigma$:
\beq
S_\sigma = {\cal S}\exp\left\{\int\limits_\sigma^\infty
d\sigma^\prime T_{\sigma^\prime} \right\},
\eeq
where ${\cal S}$ puts
operators in order of increasing scale. Then the
similarity transformation is given by the differential
``flow'' equation
\beq
\frac{dH_\sigma}{d\sigma} = -[T_\sigma,H_\sigma]
\eeq
with the boundary condition $H_\infty=H$.
Now let the eigenstates of $H_0$ be $|i\rangle$,
with $H_0|i\rangle = E_i|i\rangle$;
and write the matrix element of an operator
${\cal O}$ as ${\cal O}_{ij}=\langle i|{\cal O}|j\rangle$.
We wish to choose $T_\sigma$ in such a way that
\beq
\label{eq:Hij}
H_{\sigma ij} = f\!\left({\textstyle 
\frac{E_i-E_j}{\sigma}}\right)
G_{\sigma ij},
\eeq
where the ``similarity form factor''
$f(x)$ is a smooth function such that
\bea
\label{eq:prop}
(i)&\quad& f(0)=1;\\
\nonumber
(ii)&\quad& f(x)\to0 \quad{\rm as}\quad x\to\infty;\\
\nonumber
(iii)&\quad& f^*(x)=f(-x); 
\eea
and $G_\sigma$ is a finite hermitian operator. The similarity
function $f$ has often been chosen to be a step function in
previous work. However, this introduces nonanalyticities and in
general should be avoided.

It is convenient to first write $H_\sigma = H_0
+ H_\sigma^{int}$ and transform to the
interaction picture ${\cal O}_I(t)=e^{iH_0t}{\cal O}e^{-iH_0t}$,
so that the ``similarity flow'' equation becomes
\beq
\label{eq:deqt}
\frac{dH_{I\sigma}(t)}{d\sigma} = -i\frac{dT_{I\sigma}(t)}{dt} 
-[T_{I\sigma}(t),H_{I\sigma}(t)],
\eeq
where $H_{I\sigma}(t)=e^{iH_0t}H^{int}_\sigma e^{-iH_0t}$
--- and so the undressed interaction Hamiltonian is
$H_I(t)= \lim_{\sigma\to\infty}H_{I\sigma}(t)$. 
After Fourier transforming,
(\ref{eq:Hij}) and (\ref{eq:deqt}) become
\beq
\label{eq:fG}
H_{I\sigma}(\omega) = \sum_{ij}2\pi\delta(\omega-E_i+E_j)
H_{I\sigma ij} = 
f{\textstyle\left({\omega\over\sigma}\right)}
G_{I\sigma}(\omega)
\eeq
and
\beq
\label{eq:deqw}
\frac{dH_{I\sigma}(\omega)}{d\sigma} = \omega T_{I\sigma}(\omega) 
-\int{d\omega'\over 2\pi}
[T_{I\sigma}(\omega'),H_{I\sigma}(\omega-\omega')],
\eeq
respectively. Now we are ready to construct some suitable
similarity operators $T_\sigma(\omega)$.

\subsection{Wilson's and G{\l}azek's scheme}

Wilson and G{\l}azek first introduced similarity renormalization in the
context of light-front Hamiltonian field theory.\cite{GW}
 Their transformation
was later elaborated upon in Ref. [3]. Essentially, their choice 
of $T_\sigma$ gives 
\beq
\label{eq:Twg}
T_{I\sigma}(\omega)= {d\,\over d\sigma}\left\{
\frac{f\!\left({\omega\over\sigma}\right)-1}
{\omega f\!\left({\omega\over\sigma}\right)}
 H_{I\sigma}(\omega)\right\},
\eeq
which shall be called the WG scheme henceforth.\footnote[2]{
In the referred work the argument of $f$ was
$\frac{E_i-E_j}{E_i+E_j+\sigma}$. It is the modification to
simple energy differences used here that makes  the use of
the interaction picture convenient.} 
Note that property $(\ref{eq:prop}.i)$ of $f$ ensures that
$T_{I\sigma}(\omega)$ is finite as $\omega\to 0$, thereby
avoiding possible problems with small energy denomenators.
This is a great advantage of the similarity renormalization
scheme.
The flow equation for the choice (\ref{eq:Twg}) is
\beq
\label{eq:Hwg}
H_{I\sigma}(\omega) = f\!\left({\textstyle{\omega\over\sigma}}\right) 
\left\{ H_I(\omega) + 
\int\limits^\infty_\sigma d\sigma' \int {d\omega'\over 2\pi}
\left[ {d\,\over d\sigma'}\left\{
\frac{f\left({\omega'\over\sigma'}\right)-1}
{\omega f\left({\omega'\over\sigma'}\right)}
 H_{I\sigma'}(\omega')\right\},
 H_{I\sigma'}(\omega-\omega') \right]\right\}
\eeq
The term $d_{\sigma'} H_{I\sigma'}(\omega')$ on the righthand side
means that when we express this as a pure integral equation
it will contain terms of all orders in $H_{I\sigma}$.
Presumably, this makes the equation rather difficult to
solve.

\subsection{Wegner's scheme}

Wegner introduced his Hamiltonian flow equation to study problems
in condensed matter physics.\cite{We} In the present
notation, his equation is
\beq
\frac{dH_\sigma}{d\sigma} 
= -{1\over\sigma^3}\left[ \left[H_0,H_\sigma\right],
H_\sigma\right].
\eeq
This is a similarity transformation with the choice
$T_\sigma = [H_0,H_\sigma]/\sigma^3$. Actually, Wegner advocates
that $H_0$ include not just the free Hamiltonian but also the
number conserving part of the undressed --- or even better, the dressed
--- interacting Hamiltonian. These are certainly choices which
one can (and, perhaps, should)
make in any similarity scheme, although
they shall not be pursued further in this work.

Wegner's flow equation can be seen as resulting from a specific 
choice of the similarity function $f$ in a more general scheme.
This scheme, referred to hereafter as the W scheme,
results from the following choice of the
infinitesmal transformation operator:
\beq
\label{eq:Tw}
T_{I\sigma}(\omega) = {d\,\over d\sigma}
\left\{{\ln f\!\left({\omega\over\sigma}\right)\over\omega}\right\}
H_{I\sigma}(\omega),
\eeq
which gives the following similarity flow equation:
\beq
\label{eq:Hw}
H_{I\sigma}(\omega) = f\!\left({\textstyle{\omega\over\sigma}}\right) 
\left\{ H_I(\omega) + 
\int\limits^\infty_\sigma d\sigma' \int {d\omega'\over 2\pi}
{d_{\sigma'}\ln
f\!\left({\omega'\over\sigma'}\right)\over
\omega'f\!\left({\omega\over\sigma'}\right)}
\left[H_{I\sigma'}(\omega'),
H_{I\sigma'}(\omega-\omega') \right]\right\}.
\eeq
This closed integral equation looks more promising than
(\ref{eq:Hwg}), although it is still nonlinear in the
dressed Hamiltonian.
To recover Wegner's flow equation, then, one just choses
$f(x)=e^{-\half x^2}$, which fulfills all the
requirements (\ref{eq:prop}) of a similarity function.
Using (\ref{eq:fG}), the flow equation (\ref{eq:Hw})
becomes
\beq
\label{eq:Gw}
G_{I\sigma}(\omega) =  H_I(\omega) - 
\int\limits^\infty_\sigma {d\sigma'\over \sigma'^2}
\int {d\omega'\over 2\pi}
{F_\sigma(\half\omega+\omega',\half\omega-\omega') 
\over \half\omega + \omega'}
\left[G_{I\sigma'}({\textstyle\half}\omega+\omega'),
G_{I\sigma'}({\textstyle\half}\omega-\omega') \right],
\eeq
where
\beq
F_\sigma(\omega_1,\omega_2) =
{f\!\left({\omega_1\over \sigma}\right) 
f'\!\left({\omega_2\over \sigma}\right)\over
f\!\left({\omega_1+\omega_2\over\sigma}\right)}
\eeq
With Wegner's choice of $f$, the form factor in (\ref{eq:Gw})
is $F_{\sigma'}= (\half\omega + \omega)^2
e^{\omega^2/4\sigma'^2}
e^{-\omega'^2/2\sigma'^2}$;
and the divergent dependence on $\omega$, while
tamed by the overall factor $f({\omega\over\sigma})$
in (\ref{eq:Hw}), will nevertheless favor higher
orders in an iterative solution of $H_{I\sigma}(\omega)$.

A similarity function $f(x)$ that falls off less sharply ---
for example, of the form $(1+x^2)^{-m}$, which
goes as $x^{-2m}$ when $x\to\infty$ --- is likely to
be preferred over Wegner's choice. In particular, for
$m=1$ this latter choice gives
\beq
\label{eq:Fx2}
F_{\sigma'}({\textstyle\half}\omega+\omega',
{\textstyle\half}\omega-\omega') = -
\frac{\sigma'^3(\omega+2\omega')(\sigma'^2+\omega^2)}
{\left(\sigma'^2+(\half\omega-\omega')^2\right)
\left(\sigma'^2+(\half\omega+\omega')^2\right)^2},
\eeq
which falls off as $\omega^{-3}$ when $\omega\to\infty$.
This damps far-off-diagonal contributions to
$G_{I\sigma}(\omega)$ that come from the nonlinear
term in (\ref{eq:Gw}).
We shall see other reasons to prefer this choice in
the following.

\section{Perturbative Solution and Intermediate representation}

\subsection{In the WG and W schemes}

One may look for solutions to the similarity flow equations by
expanding in the undressed interaction $H_I(\omega)$ via
\beq
\label{eq:Gper}
G_{I\sigma}(\omega) = \sum_{n=0}^\infty
\int{d\omega_1\over 2\pi}\cdots\int{d\omega_n\over 2\pi}
g^{(n)}_{\sigma 1...n}(\omega)
H_I(\omega_1)\cdots
H_I(\omega_n)H_I(\omega-\omega_1-...-\omega_n).
\eeq
Then the functions $g^{(n)}_{\sigma k...l}(\omega)\equiv g^{(n)}_\sigma
(\omega;\omega_k,...,\omega_l), n\ge 1,$ satisfy the recursion
equations
\bea
 g^{(n)}_{\sigma 1...n}(\omega)
&=& \sum_{m=1}^{n} \int\limits_\sigma^\infty d\sigma'
\left\{ f({\textstyle{\omega_m\over\sigma'}})
g^{(m-1)}_{\sigma' 1...m-1}(\omega_m)
{d\,\over d\sigma'}\left[{1-f({\omega-\omega_m\over\sigma'})
\over\omega-\omega_m}g^{(n-m)}_{\sigma' m+1...n}(\omega-\omega_m)\right]
\right. \\
\nonumber && \quad -
\left. {d\,\over d\sigma'}\left[{1-f({\omega_m\over\sigma'})
\over\omega_m}g^{(m-1)}_{\sigma' 1...m-1}(\omega_m)\right]
{\textstyle f({\omega-\omega_m\over\sigma'})}
g^{(n-m)}_{\sigma'm+1...n}(\omega-\omega_m)
\right\}
\eea
for the WG scheme and
\bea
 g^{(n)}_{\sigma 1...n}(\omega)
&=& \sum_{m=1}^{n} \int\limits_\sigma^\infty d\sigma'
\left\{ 
{F_{\sigma'}(\omega_m,\omega-\omega_m)\over \omega_m}
-{F_{\sigma'}(\omega-\omega_m,\omega_m)\over 
\omega-\omega_m}
\right\} \\
\nonumber && \qquad\qquad\qquad \times
g^{(m-1)}_{\sigma'1...m-1}(\omega_m)\,
g^{(n-m)}_{\sigma'm+1...n}(\omega-\omega_{m})
\eea
for the W scheme, with $g^{(0)}_\sigma=1$ for both schemes.

The similarity transformed Hamiltonian is then obtained
by normal ordering the field operators in the righthand side of
(\ref{eq:Gper}), giving a unique representation in terms
of normal-ordered products of field operators:
\beq
\label{eq:Gnor}
G_{I\sigma}(\omega) = \sum_{m=0}^\infty 
\left\{ \prod_{i=1}^m
\int{d\omega_i\over 2\pi}\right\}
{\bar g}^{(m)}_{\sigma 1...m}(\omega)
:H_I(\omega_1)\cdot\cdot\cdot H_I(\omega_m)
H_I(\omega-\omega_1-...-\omega_m):.
\eeq
A given normal-ordered term with coefficient
${\bar g}^{(m)}_{\sigma}$ in the
dressed Hamiltonian will have contributions from an
infinite number of terms with
$n>m$ in the expansion (\ref{eq:Gper}), with each
contribution consisting of an integration over the product of
$g^{(n)}_{\sigma}$ with one or more functions coming from the
contractions of the field operators in (\ref{eq:Gper}).
Since these contractions do not come from time-ordered products,
one has many more contraction terms than in Feynman perturbation
theory. Indeed, the
resulting diagrammatic expansion\cite{WWH} is that of old-fashioned
perturbation theory. The output dressed Hamiltonian is
still to be solved, of course; but now we have a variety of nonperturbative
methods, such as variational techniques, that can be applied.

\subsection{Dyson's scheme}

If the scale $\sigma$ is large enough, then it is reasonable
to solve the similarity flow equations perturbatively as above;
for then only high energy processes will dress $H_{I\sigma}$,
and we have assumed for our theory that the interactions governing
such processes are weak. However, the proliferation of
diagrams in old-fashioned perturbation theory makes the above
schemes quite tedious. Even second-order calculations can be
rough; and in light-front field theory, where similarity
renormalization methods
have been in use for several years now, little (if anything)
has been accomplished at higher orders. Thus it is more than
just interesting to note that Dyson long ago developed a
similarity renormalization scheme which is based upon the
evaluation of off-energy-shell Feynman diagrams. 
Since Dyson's intermediate representation was developed over several long
papers, we can do it little justice here: for 
more details and the justification of many claims 
one should refer to the original work.\cite{D}
We shall oversimplify the presentation, change
the notation, and --- to
a certain extent --- alter the scheme itself,
in fitting it into the present work. 

Dyson considered the transformation 
${\cal O}_D(t) = S_D^{-1}(t){\cal O}_I(t)S_D(t)$
of interaction representation operators to another
representation induced by the unitary
operator
\beq
\label{eq:S_D}
S_D(t) = e^{iH_0t}S_D e^{-iH_0t} = {\cal T}\left\{
e^{-i\int_{-\infty}^t dt' H_I(t,t')}\right\}, 
\eeq
where ${\cal T}$ is the usual time-ordering operator
(acting only on the integrated times in the expansion
of the exponential) and
\beq
\label{eq:Htt'}
H_I(t,t') = g(t-t')H_I(t')
\eeq
describes a gradual switching on of the 
interaction: one requires
$g(t-t')\to 0$ as $t'\to -\infty$ and $g(0)=1$.

To be precise, Dyson introduced the function
$g$ by rescaling each occurence of the renormalized
coupling $e_R$ in $H_I(t_i)$, including the
counterterms, via $e_R\to g(t-t_i)e_R$, and included in
$H_I(t,t')$ terms depending on time derivatives of
the $g(t-t_i)$. Our modification to (\ref{eq:Htt'})
simplifies the presentation, 
but we should keep in mind that it is desirable to
take terms of order $(e_R)^n$ in $H_I(t,t')$ to be proportional
to $g^n(t-t')$, so that all terms in $H_D$ of a given
order in $e_R$ have the same strength. The modification to
our presentation needed to meet this demand is straightforward:
one would have
$H_I(t,t') = \sum\limits_{n=0}^\infty a_n g^n(t-t') H^{(n)}_I(t')$,
where $H^{(n)}_I$ is of order $(e_R)^n$ and the $a_n$ are 
dimensionless constants.

With $S_D$ and $g$ chosen as above, then,
low-energy processes, corresponding to large times,
do not modify ${\cal O}_D(t)$; whereas high-energy
processes do.
Indeed, since the function $g$ is dimensionless, it must
introduce an energy scale, which we can write explicitly
via $g(t)\to g(\sigma t)$. Dyson chose the form
\beq
\label{eq:gparam}
g(\sigma t) = \int\limits\limits_0^\infty d\Gamma G(\Gamma) 
e^{-\Gamma\sigma t}
\eeq 
(so $g$ is the Laplace transform of $G$)
and showed that if one imposes the condition $g'(0)=0$ ---
that is, $\int d\Gamma \Gamma G(\Gamma) = 0$ --- then the 
(Fourier-transformed) intermediate representation Hamiltonian obeys
\beq
H_D(\omega) \propto {1\over \omega^2} \quad {\rm as}
\quad \omega\to\infty,
\eeq
demonstrating that $S_D(t)$ induces a similarity transformation
as defined above. Note that as $\sigma\to 0$, $g\to 1$ (which
implies $\int d\Gamma G(\Gamma)=1$); and so in this limit $S_D(t)$ 
becomes the familiar time-evolution operator that
transforms from the Heisenberg to the interaction pictures.
In this sense, then, for finite $\sigma$, $S_D(t)$ 
can be considered as transforming
to an intermediate representation, which we have labelled with
a $D$ (Dyson picture) since the label $I$ has already been 
taken.
The intermediate Hamiltonian is
\bea
\label{eq:Hd0}
H_D(t) &=& S_D^{-1}(t)[H_0+H_I(t)] S_D(t) \\
\nonumber 
&=& \sum_{n=0}^\infty i^n
\int\limits_{-\infty}^t dt_1\, g\left(\sigma(t-t_1)\right)
\int\limits_{-\infty}^{t_1} dt_2\, g\left(\sigma(t-t_2)\right)
\cdots \\
\nonumber && \qquad \cdots
\int\limits_{-\infty}^{t_{n-1}} dt_n\, g\left(\sigma(t-t_n)\right)
[ H_I(t_n),... [ H_I(t_2),[H_I(t_1),H_0 + H_I(t)]]...],
\eea
and the infinitesmal similarity transformation operator
associated with $S_D(t)$ is
\bea
T_D(t) &=& S_D^{-1}(t){d\,\over d\sigma} S_D(t) \\
\nonumber 
&=& -\sum_{n=1}^\infty i^n
\int\limits_{-\infty}^t dt_1\, d_\sigma g\left(\sigma(t-t_1)\right)
\int\limits_{-\infty}^{t_1} dt_2\, g\left(\sigma(t-t_2)\right)
\cdots\\
\nonumber && \qquad \cdots
\int\limits_{-\infty}^{t_{n-1}} dt_n\, g\left(\sigma(t-t_n)\right)
[ H_I(t_n),... [ H_I(t_2),H_I(t_1)]...].
\eea
Thus Dyson's transformation is an explicit perturbative
representation of a similarity transformation, and the label D will
always imply a $\sigma$ scale dependence.

Writing $H_D(t)=H_0 + H^D_I(t)$ and integrating by parts,
we may express (\ref{eq:Hd0}) as
\bea
\label{eq:Hd1}
H^D_I(t) &=& \sum_{n=0}^\infty i^n
\int\limits_{-\infty}^t dt_0\, d_{t_0}g\left(\sigma(t-t_0)\right)
\int\limits_{-\infty}^{t_0} dt_1\, g\left(\sigma(t-t_1)\right)
\int\limits_{-\infty}^{t_1} dt_2\, g\left(\sigma(t-t_2)\right)
\cdots \\
\nonumber && \qquad \cdots
\int\limits_{-\infty}^{t_{n-1}} dt_n\, g\left(\sigma(t-t_n)\right)
[ H_I(t_n),... [ H_I(t_2),[H_I(t_1),H_I(t_0)]]...].
\eea
Note that for finite $\sigma$ the $t$-dependence of the intermediate
Hamiltonian is still that of an interaction picture operator:
$H^D_I(t)=e^{iH_0t}H^D_Ie^{-iH_0t}$.
Now (\ref{eq:Hd1}) may be expressed in
the alternate form
\bea
\label{eq:Hd2}
H^D_I(t) &=& -\sigma
\int\limits_{-\infty}^t dt_0\, g'\left(\sigma(t-t_0)\right)\\
\nonumber && \qquad \times
\sum_{m,n=0}^\infty {i^m\over m!}{(-i)^n\over n!}
\int\limits_{-\infty}^\infty dt_1\, g\left(\sigma(t-t_1)\right)
\cdots
\int\limits_{-\infty}^\infty dt_{m+n}\, g\left(\sigma(t-t_{m+n})\right)\\
\nonumber && \qquad\qquad\quad \times
\overline{\cal T}\left\{ H_I(t_{n+1})... H_I(t_{n+m})\right\}
{\cal T}\left\{ H_I(t_0)... H_I(t_n)\right\},
\eea
where $\overline{\cal T}$ orders in the sense opposite to
${\cal T}$. This latter form allows the normal-ordered
Hamiltonian to be built up in terms of connected Feynman
graphs. Let us briefly review Dyson's derivation.

First, as above, we must express the intermediate Hamiltonian in terms
of a sum over normal-ordered products of field operators.
For the chronologically ordered and antichronologically ordered time
products 
(the latter results from the Hermitian conjugate of the former)
we can use Wick's theorem.
Field contractions from these are then the usual Feynman propagators
(and their complex conjugates), and we are left with a product of
two sets of normal-ordered operators. In normal ordering this 
remaining product, whose operators are not related chronologically,
one finds field contractions that do not correspond to Feynman
propagators. For example, we have
\beq
\psi_\alpha(x)\bar\psi_\beta(x') = \;\,
:\psi_\alpha(x)\bar\psi_\beta(x'): + \; S^{\alpha\beta}_+(x-x'),
\eeq
where the ``abnormal'' contraction $S^{\alpha\beta}_+(x-x')$ 
is specified below.
The products $\psi_\beta(x)\bar\psi_\alpha(x')$ and 
$A_\mu(x)A_\nu(x')$ are expanded
similarly, giving the abnormal contractions 
$S^{\alpha\beta}_-
(x-x')$ and $\delta_{\mu\nu}D_+(x-x')$, 
respectively. We have
\beq
D_+(x-x') = \int \widetilde{dk}_B
e^{-ik\cdot(x-x')},
\eeq
and
\beq
iS^{\alpha\beta}_\pm(x-x') =  
\int \widetilde{dk}_F
\left(\not\! k \pm \! m_F\right)_{\alpha\beta}
e^{-ik\cdot(x-x')},
\eeq
where the invariant phase space measure is
\beq
\widetilde{dk}_{B,F} =
{d^4k\over(2\pi)^3}\theta(k_0)\delta(k^2-m^2_{B,F})
\eeq
with $m_B$ and $m_F$ are the 
boson and fermion masses.

Now, using Dyson's parameterization (\ref{eq:gparam}), 
we find
\beq
\label{eq:Hd3}
H^D_I(\omega) = \sigma\sum_{m,n=0}^\infty
{i^m\over m!}{(-i)^n\over n!} \left\{ \prod_{i=0}^{m+n}
\int\limits_0^\infty d\Gamma_i G(\Gamma_i)\right\}
\sum_{\{p\}}
{\cal M}^{\{p\}}_{mn}(\omega,\Gamma) 
{\cal N}_{\{p\}}^{{\bar F}FB},
\eeq
where the latter sum represents
\beq
\sum_{\{p\}} = \sum_{{\bar F},F,B=0}^\infty 
\left\{ 
\prod_{i=1}^{{\bar F}+F+B} 
\int {d^4p_i\over (2\pi)^4} \right\}
\eeq
and ${\cal N}_{\{p\}}^{{\bar F}FB}$ is the normal-ordered product of 
fields
\beq
\label{eq:np}
{\cal N}^{{\bar F}FB}_{\{p\}} =\, :\! 
\psi^\dagger(p_1)\cdots\psi^\dagger(p_{\bar F})\psi(p_{{\bar F}+1})
\cdots \psi(p_{{\bar F}+F})
A(p_{{\bar F}+F+1})\cdots A(p_{{\bar F}+F+B})\! :\, ,
\eeq
with 
\bea
\psi(p) &=& \int d^4x e^{ip\cdot x}\psi(x)\\
 &=&  \nonumber 
2\pi \delta(p^2-m_F^2)\sum_\alpha \left\{ \theta(p^0)
u^{(\alpha)}(p) b_\alpha(p) + \theta(-p^0)
v^{(\alpha)}(-p) d^\dagger_\alpha(-p)\right\},
\eea
$\psi^\dagger(p) = \int d^4x e^{-ip\cdot x}\psi^\dagger(x)$
and $A(p)=\int d^4x e^{ip\cdot x} A(x) = A^\dagger(-p)$. 
Evidently, we have
\bea
\label{eq:expand}
&&\sum_{\{p\}} {\cal M}^{\{p\}}_{mn}\, {\cal N}_{\{p\}}
= \left\{ \prod_{i=0}^{m+n}
\int\limits_{-\infty}^\infty dt_i \right\}
\Gamma_0\int\limits_{-\infty}^\infty dt\, 
e^{-i(\omega-\sigma\Lambda_{m+n})t}\theta(t-t_0) \\
&& \nonumber \qquad\qquad \times \overline{\cal T}\left\{
e^{\sigma\Gamma_{n+m} t_{n+m}}H_I(t_{n+m})\cdots
e^{\sigma\Gamma_{n+1} t_{n+1}}H_I(t_{n+1})\right\}
{\cal T}\left\{
e^{\sigma\Gamma_0 t_0}H_I(t_0)\cdots
e^{\sigma\Gamma_n t_q}H_I(t_n)
\right\}
\eea
where $\Lambda_k = i(\Gamma_0 + \Gamma_1 + ... + \Gamma_k)$.
Assume that $m\ge 1$. Dyson showed that in the diagrammatic expansion
based upon (\ref{eq:expand}) only connected graphs contribute,
and therefore ${\cal M}_{mn}^{\{p\}}$ must contain some $L \ge 1$
integrations over momenta $k_l$ corresponding to ``abnormal''
contractions between operators at times $t_i$, $i> n$ and
times $t_j$, $j\le n$. From the definitions above, we see
that these abnormal contractions will contribute
Fourier components $e^{-ik_l\cdot(x_i-x_j)}$ with the
zeroth component $k_l^0\ge 0.$ We have then $k_l^0\ge \min(m_B,m_F)$
for all $L$ abnormal contractions.

Next we make the shift of integration variables
$t_j\to t_j + t$ for $j\le n$ and consider 
${\cal M}_{mn}^{\{p\}}$ as a function of 
the variables $\Gamma_i$ and $m_{B,F}$. In calculating 
${\cal M}$, we consider the $\Gamma_i$ to be pure imaginary
numbers in order to perform the time integrations, which
have been extended to positive infinity. 
After these changes, the integration over $t$ in
(\ref{eq:expand}) yields a delta-function
\beq
\label{eq:delta}
{\cal M}_{m,n}^{\{p\}}(\omega,\Gamma)
\propto \delta(
\omega - i\sigma\sum_{i=n+1}^{n+m}\Gamma_i -
\sum_{j=0}^{n}\omega_j -
\sum_{l=1}^N k^0_l),
\eeq
where $\omega_j$ is the algebraic sum of the zeroth components
of the external momenta $p_k$ that meet vertex $j$, so that
$\omega = \sum\limits_{k=1}^{\bar F} p_k^0 -
\sum\limits_{k=1}^{{\bar F}+F+B} p_k^0 = 
\sum\limits_{i=0}^{m+n}\omega_i$.
If we assume that $m_B$ and $m_F$ are large enough,
then
\beq
\label{eq:cond}
\sum_{l=1}^L k^0_l \ge L\min(m_B,m_F) > \omega
-i \sigma \sum_{i=n+1}^{n+M} \Gamma_i - \sum_{j=0}^n \omega_j
\eeq 
so that (\ref{eq:delta}) cannot be satisfied and
${\cal M}^{\{p\}}_{mn}$ vanishes for $m\ge 1$. Thus
for ${\rm Re}(\Gamma_i)=0$ and $m_{B,F}$ satisfying
(\ref{eq:cond}), ${\cal M}$
may be found from (\ref{eq:expand}) with $m=0$, which then
is simply an off-energy-shell Feynman amplitude.
Dyson argued that considered as a complex function of
the variables $m_{B,F}$ and $\Gamma_i$, ${\cal M}$ is
analytic, and thus this representation
of ${\cal M}$ is unique. So to obtain
the intermediate Hamiltonian, one need simply
calculate ${\cal M}$ with the variables constrained
as above (so that one can use generalized Feynman
graphs), then analytically continue first
$m_{B,F}$ from large values down to their
physical values and then the $\Gamma_i$ from
pure imaginary back to pure real numbers (in practice,
this analytic continuation is trivial). Finally,
the integrals over the $\Gamma_i$ in (\ref{eq:Hd3})
may be performed.

\subsection{Doubled Feynman graphs and spurions}

Dyson described a graphical 
method,\footnote[2]{Although, sadly, his papers have no
figures.}
using what he called ``doubled'' Feynman diagrams,
for calculating intermediate representation operators
from expressions such as (\ref{eq:expand}).
A doubled Feynman diagram
is an ordinary connected $n+1$-point Feynman graph
in which a simply connected ``tree'' of (fermion and/or
boson) propagator lines are drawn double. An example
is given in Figure 1. The doubled lines account for
the factors $e^{\sigma\Gamma_i t_i}$ that occur at each point
$x_i$ and spoil energy conservation. If it is possible to
draw more than one 
doubled Feynman graph from a given Feynman graph, one
of these is chosen arbitrarily and the rest are ignored
(they are not distinct, each one being obtained from any
other by a suitable change in the internal momentum
variables). Dyson then describes in detail how to shift
the momenta of the doubled propagators consistently to
account for the dependence of ${\cal M}$ upon the $\Gamma_i$.
The rules are somewhat involved.

We instead outline here an alternate but entirely
equivalent (and hopefully more transparent) diagrammatic scheme
that borrows from a formulation of perturbation theory due
to Kadyshevsky.\cite{K,Karma} 
To do this, we
express the step function in (\ref{eq:expand})
as a complex integration through the familiar representation
\beq
\theta(t-t_0) = \int {d\nu\over 2\pi i}
{e^{i\nu(t-t_0)}\over \nu-i\varepsilon} 
= \int {d\nu\over 2\pi i}
{e^{i\nu\sigma(t-t_0)}\over \nu-i\varepsilon};
\eeq
define the four-vector $\lambda_\mu = \sigma n_\mu$,
where $n^2=1$ and $n\cdot x = x^0$; and write
$H_I(t_i)=\int d^3x_i{\cal H}_I(x_i)$.
Then (\ref{eq:expand}) becomes
\bea
\label{eq:ex2}
\sum_{\{p\}} {\cal M}^{\{p\}}_{mn}\, {\cal N}_{\{p\}}
&=& \delta_{m0}
\int{d\nu\over 2\pi}{-i\Gamma_0 \over \nu -i\varepsilon}
\int dt\, e^{-i(\omega-\sigma\Lambda_n-\sigma\nu)t}
\left\{ \prod_{i=0}^q\int d^4x_i \right\}\\
&& \nonumber \qquad \times
{\cal T}\left\{
e^{(\Gamma_0-i\nu)\lambda \cdot x_0}{\cal H}_I(x_0)
e^{\Gamma_1\lambda \cdot x_1}{\cal H}_I(x_1)\cdots
e^{\Gamma_n\lambda\cdot x_n}{\cal H}_I(x_n)
\right\}.
\eea
From (\ref{eq:Hd3}) and (\ref{eq:ex2}), we find the
Fourier transform of the interaction part of the
total intermediate representation momentum is
\bea
\label{eq:Hd4}
{\cal P}_{I\mu}^D(p)
&=& n_\mu H_I^D(p) = \lambda_\mu \sum_{n=0}^\infty
{(-i)^n\over n!} \left\{ \prod_{i=0}^n \int d\Gamma_i
G(\Gamma_i)\right\}
\int{d\nu\over 2\pi}{-i\Gamma_0 \over \nu -i\varepsilon}\\
&& \nonumber \qquad\quad\times
\int d^4x e^{-i(p-(\Lambda_n+\nu)\lambda)\cdot x}
\left\{ \prod_{i=0}^n\int d^4x_i \right\}\\
&& \nonumber \qquad \times
{\cal T}\left\{
e^{(\Gamma_0-i\nu)\lambda \cdot x_0}{\cal H}_I(x_0)
e^{\Gamma_1\lambda \cdot x_1}{\cal H}_I(x_1)\cdots
e^{\Gamma_n\lambda\cdot x_n}{\cal H}_I(x_n)
\right\}.
\eea
We expand the time-ordered product in terms of operator products
(\ref{eq:np}) and perform the configuration space integrals
in (\ref{eq:Hd4}), so that the  resulting coefficient 
${\cal M}^{\{p\}}_n$ is a
function of $\nu$, the $n+1$ variables $\Gamma_i$, 
the $N$ external momenta $p_j$, and the total momentum
difference $p$. These amplitudes will consist of
integrations of $m\ge n$ Feynman propagators depending
on $m$ internal momenta $q_k$.
Shifting the space-time variables in (\ref{eq:Hd4})
as $x_i\to x_i + x$, we see that
${\cal M}$ is multiplied by an overall $\delta$-function
\beq
{\cal M}^{\{p\}}_n(p,\{\Gamma_i\},\nu) = (2\pi)^4
\delta^{(4)}(p-p^O+p^I) 
{\cal M}^{
\{p\}}_n(\{\Gamma\},\nu).
\eeq
This arises from integration over $x$. Here $p^O$ is
the sum of outgoing and $p^I$ the sum of ingoing externa 
momenta. The remaining
integrals over the ${x_i}$ also give $\delta$-functions,
but the factors $\Gamma_i$ and $\nu$ spoil energy conservation
at the vertices. For each vertex $x_i$ we get a function
\beq
(2\pi)^4
\delta^{(4)}(p^O_i - p^I_i + q^O_i - q^I_i
 - (i\Gamma_i + \delta_{i0}\nu)\lambda)
\eeq
where $p^O_i$ is sum of the external
momenta leaving $x_i$, $q_i^I$ is the sum of the internal
momenta entering $x_i$, and so forth.

The amplitudes ${\cal M}^{\{p\}}_n$ are connected $n+1$-point
Feynman diagrams whose dependence upon $\nu$ and the $\Lambda_i$
may be accounted for by drawing an additional 
directed dashed line as in Figure 2, which is equivalent to Figure 1.
This is very similar to the spurion line of
Kadyshevsky's diagrammatic scheme.\cite{K} Although the rules for our
scheme are quite different than those of Kadyshevsky's, we shall
still call our dashed line a spurion line. We treat the variables
$\Lambda^\mu = -\lambda^\mu\nu$ and 
$\Lambda_i^\mu = i\lambda^\mu\sum_{j=0}^i \Gamma_j$ as momenta and
stipulate that the spurion line carries momentum as shown
in Figure 2 --- namely, the spurion line entering vertex $i$
has momentum $\Lambda_i$, and that exiting vertex $0$ has
momentum $\Lambda$. The spurion line touches each vertex
exactly once, proceeding in decreasing order from vertex $n$ to  
$0$. A relabeling of the vertices 1 through $n$ will not
produce a distinct diagram since (\ref{eq:Hd4}) is symmetric
with respect to $\Gamma_i \leftrightarrow \Gamma_j$, $i\ne j
\ne 0$, and so for a given choice of vertex 0 we may choose
any ordering of 1 through $n$ and ignore all others, for they
just sum to cancel the factor ${1\over n!}$ in (\ref{eq:Hd4}).
We may think of the spurion line as ordering the different
vertices in time; however, it is only the order of vertex 0
that need be distinguished from that of the others.

With the spurion line considered as carrying
momentum in the way described above, 
we have the usual four-momentum conservation
at each vertex; and the evaluation of the amplitude ${\cal M}$
proceeds as that of an ordinary Feynman graph. The spurion line
serves only to shift the internal momenta so that the dependence
on the $\Lambda_i$ is now carried by the Feynman
propagators --- this is precisely the role of the doubled lines in Dyson's 
formulation. There will also always be an overall $\delta$-function
of the form $\delta^{(4)}(p - \Lambda_n + \Lambda)$, which expresses
conservation of total three-momentum but not energy.
The dependence
of ${\cal M}$ upon the variables $\Gamma_0$ and $\nu$
comes solely from this latter $\delta$-function, and so the
integrals over these variables in (\ref{eq:Hd4}) will yield the
similarity function described earlier, a function that by choice
of $G(\Gamma_0)$ falls off at least as rapidly as ${1\over \omega^2}$
when the total energy difference $\omega=p^0$ diverges.

The most obvious advantage of Dyson's scheme is the apparent
invariance of the expression (\ref{eq:Hd4}). The $\Gamma_i$ and
$\nu$ are dimensionless scalars, and covariance is broken only
by $\lambda = \sigma n$, which was chosen to have a special form. 
We note now that the choice $n = (1,0,0,0)$ above gives
the equal-time formulation described here, but one could just
as well choose $n=n_+$ with $n^2_+=0$, $n_+\cdot x = x^+
= x^0 + x^3$. This latter choice
of $\lambda$ ensures that (\ref{eq:Hd4}) yields the light-front
formulation of the Hamiltonian, with $x^+$ the ``light-front time.'' 
It is important, however, that divergences arising from the
integrations 
in (\ref{eq:Hd4}) are independent of $\lambda$, for they
occur when the internal momenta of our Feynman graphs
diverge with respect to the variables $\Lambda_i$. Thus the
determination of the counterterm structure of $H_I$ is entirely
equivalent to ordinary Feynman perturbation theory.\footnote[2]{As
with Feynman perturbation theory, the light-cone gauge choice
$A^+ = n_+\cdot A =0$ is noncovariant and necessitates the
introduction of counterterms that depend on $n_\pm=(1,0,0,\pm 1)$.
Such counterterms, of course, cannot depend on $\sigma$.}
In particular, the usual rules for
power counting, separation of divergences and
multiplicative renormalizability apply, 
results that are by no means clear in other schemes for light-front
variables.

\subsection{An example}

As an example we shall look at the quark-antiquark potential
in light-front QCD. We do not discuss at length
any of the many interesting, enticing and
puzzling features of light-front QCD. For more details see
Ref. [3] and the many references listed therein.
Here we shall only outline the calculation and give the results. 
When formulated using light-front variables\footnote[3]{Light-front
longitudinal momentum is $p^+ = n_+\cdot p = p^0 + p^3$, light-front
energy is $p^- = n_-\cdot p = p^0-p^3$, and transverse
momentum is $p_\perp^i = p^i$, $i=1,2$. The scalar product
is $p\cdot x = \half p^+x^- + \half p^- x^+ - p_\perp\cdot x_\perp$,
where the notation $p_\perp\cdot x_\perp$ means
$p^1 x^1 + p^2 x^2$. Occasionally, it convenient to define the
light-front 3-vector $\bar{p} = (p^+,p_\perp)$ and the product
$\bar{p}\cdot\bar{x} = p^+ x^- - p_\perp\cdot x_\perp$.
On-shell particles satisfy $p^- = \frac{p_\perp^2 + m^2}{p^+}$,
from which it follows that transverse momenta scale as masses but
longitudinal momenta scale differently.}
and light-cone gauge, 
half the degrees of freedom of QCD are constrained
and one may develop an effective theory in which the
gluons and quarks are two-component fields.\cite{ZH} For some calculations
it is more appropriate to use the original four-component
constrained theory\cite{BDS,BNS} --- for example, to demonstrate
that the Slavnov-Taylor identities are fulfilled.\cite{LM,BNS} In general,
however, calculations are much easier in the two-component theory,
and we use just this theory here.
The Feynman rules for the two-component effective theory 
are given in the appendix of Ref. [9].  
We calculate the coefficient of the term ${\cal N}^{220}_{\{p\}}
=\, :\!\psi^\dagger(p_3)\psi^\dagger(-p_2)\psi(-p_4)\psi(p_1)\!:$ in the
intermediate Hamiltonian to second order in the undressed
coupling $g_R$.
Perry\cite{P} first showed using a discrete similarity transformation
(with the similarity function $f$ chosen to be a step function)
that one obtains a logarithmically confining potential in the
dressed Hamiltonian. Zhang\cite{Z} later showed that a continuous
similarity transformation such as those described here (specifically,
the WG scheme of Ref. [3]
with $f$ a step function) yields a potential of the
same form. Wilson and Robertson\cite{WR} gave general arguments (and
caveats) for the appearance of such a confining potential in 
similarity transformed light-front QCD Hamiltonians.
Here, we show how such a potential arises within
Dyson's scheme. 

The graphs that contribute at second order are shown in
Figure 3. Graph 3a comes from a term ${\cal H}^{(2)}_I(x)$
in the bare Hamiltonian
that arises from eliminating the constrained components of the
gluon field.\cite{ZH} This term is proportional to $(g_R)^2$, and in
the spirit of our prior discussion, we take the term
${\cal H}_I^{(2)}(x,x')$ appearing in the definition (\ref{eq:S_D})
of $S_D$ to be proportional to $g^2(\lambda\cdot(x-x'))$. The change
to our momentum space Feynman rules caused by this modification is slight:
we need only alter the way the spurion line carries the
momenta $\Gamma$. If each vertex $i$ in a given $n+1$-vertex diagram
corresponds to a term in the
Hamiltonian proportional to $(g_R)^{m_i}$, then we have 
$m = \sum_{i=0}^n m_i$ factors $\Gamma_{ij}$, $1\le j\le m_i$,
corresponding to the functions $g^{m_i}(\lambda\cdot(x-x_i))$.
Then the diagrams are exactly the same, except for the 
redefinition $\Lambda_i = i\lambda \sum_{j=0}^i\sum_{k=1}^{m_j} 
\Gamma_{j,k}$. The integration in (\ref{eq:Hd4}) is now over
\beq
\left\{\prod_{i=0}^n \prod_{j=1}^{m_i} \int d\Gamma_{ij}
G(\Gamma_{ij})\right\} \int {d\nu\over 2\pi}{m_0\Gamma_{01}
\over i(\nu - i\varepsilon)}.
\eeq
The amplitudes corresponding to the graphs in Figure 3 are thus
\beq
\label{eq:Ma}
i{\cal M}_a = -4ig^2_B T^a T^a {1\over [p^+_1-p^+_3]^2}
(2\pi)^4\delta^{(4)}(p - \Lambda_0 + \Lambda),
\eeq
where $p = p_3 + p_4 - p_2 - p_1$ and 
$\Lambda_0 = i\lambda (\Gamma_{01}+\Gamma_{02})$;
\bea
\label{eq:Mb}
i{\cal M}_b &=& \left\{ {q^i_\perp\over [q^+]}
- {\sigma_\perp\cdot p_{3\perp}-im \over 2[p^+_3]}\sigma^i_\perp
- \sigma^i_\perp{\sigma_\perp\cdot p_{1\perp}+im \over 2[p^+_1]}
\right\}
{-4ig^2_B T^a T^a\over q^2+i\varepsilon}\\
\nonumber && \quad
\left\{ {q^i_\perp\over [q^+]}
- {\sigma_\perp\cdot p_{2\perp}-im \over 2[p^+_2]}\sigma^i_\perp
- \sigma^i_\perp{\sigma_\perp\cdot p_{4\perp}+im \over 2[p^+_4]}
\right\}
(2\pi)^4\delta^{(4)}(p - \Lambda_1 + \Lambda),
\eea
where $p$ is as above, $\Lambda_0=i\lambda\Gamma_{01}$,
$\Lambda_1=i\lambda\Gamma_{11}+\Lambda_0$, and
$q=p_3-p_1+\Lambda_0-\Lambda_1$;
and, finally, ${\cal M}_c$ is obtained from ${\cal M}_b$ by the
substitutions $\Lambda\leftrightarrow -\Lambda_1$,
 and $\Lambda_0\to -\Lambda_0$. In general,
we need to regulate the singularities as the longitudinal
momentum $q^+=n\cdot q\to 0$. As is usual for Feynman diagrams
in light-front variables, we use the Mandelstam-Leibbrandt
prescription\cite{ML}
\beq
{1\over [q^+]} = {q^-\over q^+ q^- + i\varepsilon}
= {1\over q^+ + i\varepsilon \Re(q^-)}.
\eeq 
For the four-point interaction of (\ref{eq:Ma}), we interpret this
prescription to be
\beq
{2\over [p_1^+-p_3^+]^2} \to
\left\{ {p_1^--p_3^-\over (p^+_1-p^+_3)(p^-_1-p^-_3)+i\varepsilon}
\right\}^2 +
\left\{ {p_4^--p_2^-\over (p^+_1-p^+_3)(p^-_4-p^-_2)+i\varepsilon}
\right\}^2.
\eeq

The potential $V(p_1,p_2,p_3,p_4)$ is proportional to
the sum of these terms, ${\cal M}_a+{\cal M}_b+{\cal M}_c$
integrated over the variables $\Gamma_{ij}$.
We are interested in how this potential behaves at large
separations, for which the terms proportional to $[q^+]^{-1}$
will dominate the expressions in the brackets in (\ref{eq:Mb}).
Summing the three diagrams and relabelling the $\Gamma$, we have then
for the long-range part of the potential
\bea
\label{eq:Vlr}
iV_{LR}\delta^{(3)}({\bar p}) &=& -4g_R^2 T^a T^a \delta^{(3)}({\bar p})
\int\limits_0^\infty d\Gamma_0 G(\Gamma_0)
\int\limits_0^\infty d\Gamma_1 G(\Gamma_1)
{\sigma\Gamma_0 \over p^- - i\sigma(\Gamma_0 + \Gamma_1)}\\
\nonumber && \quad \times
\left\{ {(q^-)^2 \over (q^+ q^- + i\varepsilon)^2} 
\cdot{ q^+ q^-\over q^+ q^- - q_\perp^2 +i\varepsilon}
+ {(\tilde{q}^-)^2 \over (q^+\tilde{q}^- + i\varepsilon)^2}
\cdot{ q^+ \tilde{q}^-\over q^+ \tilde{q}^- - q_\perp^2 +i\varepsilon}
\right\},
\eea
where $q^- = p_3^- - p_1^- - i\sigma \Gamma_1$ and 
$\tilde{q}^- = p_2^- - p_4^- + i\sigma \Gamma_1$. For small
${\bar q} = {\bar p}_3 - {\bar p}_1 = {\bar p}_2 - {\bar p}_4$,
the energy difference $p^-$ can be ignored with respect to $\sigma$.
We calculate $V_{LR}(x^-,x_\perp)$, the Fourier transform with
respect to ${\bar q}$ of (\ref{eq:Vlr}) above under this
approximation. We need to be careful, however, for $V_{LR}(0,0)$
diverges. This is a result of well-known infrared singularities
due to the choice of light-cone gauge, singularities that are
known to cancel (in perturbation theory, at least) when evaluating
gauge invariant quantities.

Now, when using light-front variables, we need to take into account
that transverse and longitudinal variables scale differently. Thus
there is a transverse mass scale $\mu$ and a longitudinal momentum
scale $\rho$. The energy scale is then $\sigma = \mu^2/\rho$. It is
natural to use dimensional regularization to handle both ultraviolet
and infrared divergences in our scheme by taking the number of
transverse dimensions to be $d$ instead of 2. Then we must introduce
an arbitrary (transverse) mass scale, which we take to be $\mu$. This
means the similarity flow is governed by the longitudinal scale
$\rho$. That the couplings now run with two scales will
have profound implications at higher orders, but this does not
concern us at present. Setting $p^-\to 0$ in (\ref{eq:Vlr}),
generalizing to $d$ transverse dimensions 
and taking the Fourier transform, we find
\bea
\label{eq:Vlr2}
V_{LR}(x^-,x_\perp) &=& {g_R^2\mu^2\over 4\pi\rho}
T^a T^a \int\limits_0^\infty d\Gamma_0 G(\Gamma_0)
\int\limits_0^\infty d\Gamma_1 G(\Gamma_1) {4\Gamma_0\Gamma_1
\over \Gamma_0 + \Gamma_1}\\
\nonumber && \quad \times
\left\{ {2\over 2-d} + \gamma + 
E_1\left(\Gamma_1{\mu^2 x_\perp^2\over 4\rho|x^-|}\right) 
+ \ln\left({\mu^2 x_\perp^2\over 4\pi}\right)
\right\},
\eea
where $\gamma$ is the Euler-Mascheroni constant and 
$E_1(x) = -Ei(-x)$ is the exponential integral function. This
gives a long-range logarithmic potential independent of the choice
of $G(\Gamma)$, which just affects its strength. Indeed, we can
use the relation $E_1(z)\to -\gamma - \ln z$ for $z$ small
to see that
\beq
V_{LR} \sim \ln \rho|x^-|
\quad{\rm for}\quad 4\rho|x^-|\gg \Gamma_1\mu^2 x_\perp^2;
\eeq
otherwise, $E_1$ is of order 1 and the $\ln \mu^2 x_\perp^2$
term dominates. The divergent piece in (\ref{eq:Vlr2}) can
be seen to be of infrared nature since it arises
from a factor $\Gamma({d\over 2} -1)$, which is regulated
by increasing $d$.

We have dropped a short range term from the potential
in going from (\ref{eq:Vlr}) to (\ref{eq:Vlr2}), namely:
\bea
V_{SR}(x_\perp) &=& {g_R^2\mu^2 T^a T^a\over 4\pi\rho}
\int\limits_0^\infty d\Gamma_0 
\int\limits_0^\infty d\Gamma_1 
{2\Gamma_0 G(\Gamma_0)\Gamma_1 G(\Gamma_1) 
\over \Gamma_0 + \Gamma_1} 
\left\{ \rule{0cm}{.7cm}
-i\cos^{-1}({\hat x}_\perp\cdot{\hat p}_{3\perp})
\right. \\ 
\nonumber && \!\!\!
-i\cos^{-1}({\hat x}_\perp\cdot{\hat p}_{4\perp}) + 
\left.\int\limits_{|x_\perp||p_{3\perp}|}^\infty {dt\over t}
J_0(t) e^{-i{\hat x}_\perp\cdot {\hat p}_{3\perp}t} +
\int\limits_{|x_\perp||p_{4\perp}|}^\infty {dt\over t}
J_0(t) e^{i{\hat x}_\perp\cdot {\hat p}_{4\perp}t}
\right\},
\eea
The dependence on the transverse momenta $p_{3\perp}$ and
$p_{4\perp}$ in this term arises from the Mandelstam-Leibbrandt
prescription for the infrared singularities. This potential is clearly
small at large $x_\perp$; however, at small $x_\perp$ it gives
a $-\ln{\mu^2 x_\perp^2}$ contribution that precisely cancels
the logarithmic term in (\ref{eq:Vlr2}). Of course, to find the
correct short distance form one has to also include the other terms
from (\ref{eq:Mb}) and (\ref{eq:Vlr}) that were dropped
in arriving at (\ref{eq:Vlr2}). 

Finally, a simple choice of
$G$ that satisfies the above conditions 
is $G(\Gamma)=\delta(\Gamma-1)+\delta'(\Gamma-1)$. This yields
\beq
\label{eq:Vlr3}
V_{LR}(x^-,x_\perp) = {g_R^2\mu^2 T^a T^a \over 4\pi\rho}
\left\{ {2\over 2-d} + \gamma + 
E_1\left({\mu^2 x_\perp^2\over 4\rho|x^-|}\right) 
+ \ln\left({\mu^2 x_\perp^2\over 4\pi}\right)
\right\},
\eeq
where we have dropped yet another short range piece.
Since the divergent constant in (\ref{eq:Vlr3}) is infrared
in nature, it does not get cancelled by a counterterm. This is
important. Perry\cite{P} has argued that this divergence is cancelled
by a similar infrared divergence in the quark self-energy
for color singlet states, but that for color nonsinglet states
this cancellation does not occur.

\section{Toward a Nonperturbative Solution}

\subsection{Dyson-Schwingerish Equations}

Within the WG and W schemes,
there does not appear to be any great advantage in solving
for the dressed Hamiltonian by using the perturbative expansion
(\ref{eq:Gper}). Rather, we can express the dressed
Hamiltonian from the outset in terms of normal-ordered products
of field operators as follows:
\beq
\label{eq:Hnor}
H_{I\sigma}(t) = \sum_{\{p\}}
h^{\cal N}_{I\sigma}(p_1,...,p_{{\bar F}+F+B})\,
e^{ip^0 t} \,{\cal N}_{{\bar F}FB} 
\equiv \sum_{\{p\}} h_{I\sigma}^{\{p\}}
{\cal N}_{\{p\}}(t),
\eeq
where $p = p^O-p^I$, as before, and
$h_{I\sigma}^{\{p\}} = {\bar h}_{I\sigma}^{\{p\}}
(2\pi)^3\delta^{(3)}(\vec p)$ --- that is, 
total 3-momentum is conserved. 
Taking the Fourier transform
of (\ref{eq:Hw}) ---
we only consider the W scheme here ---
we find
\beq
\label{eq:Hw2}
H_{I\sigma}(t) = \int dt' \widetilde{f}(t-t')  H_I(t') + 
\int\limits^\infty_\sigma d\sigma' \int dt' dt''
\widetilde{F}_{\sigma\sigma'}(t-t'',t''-t') 
H_{I\sigma'}(t')H_{I\sigma'}(t'').
\eeq
where $\widetilde{f}(t)$ is the Fourier transform of $f(\omega)$
and
\beq
\label{eq:Ftil}
\widetilde{F}_{\sigma\sigma'}(t_1,t_2) 
= \int {d\omega\over 2\pi} {d\omega'\over 2\pi}
e^{i\omega t_1} e^{i\omega' t_2}
{f({\omega\over\sigma})\over f({\omega\over\sigma'})}
\left\{ {d_{\sigma'}\ln f({\omega'\over\sigma'})\over\omega'} -
{d_{\sigma'}\ln f({\omega-\omega'\over\sigma'})\over
\omega-\omega'} \right\}
\eeq
Now let us choose the similarity function motivated earlier, namely,
$f(x) = (1+x^2)^{-1}$. Then the term in the brackets in (\ref{eq:Ftil})
has a pole structure in $\omega'$ that yields step functions
in $t_2$ after integration. We find
\bea
\label{eq:Hw3}
H_{I\sigma}(t) &=& {\sigma\over 2}\int dt' 
e^{-\sigma|t-t'|}  H_I(t') 
- {i\sigma\over 4}
\int\limits^\infty_\sigma d\sigma' 
{\sigma'^2-\sigma^2\over\sigma'^3}
\int dt' dt'' e^{-\sigma'|t'-t''|} 
\\
\nonumber && \quad 
\times \left(
e^{-\sigma|t-t'|}+e^{-\sigma|t-t''|}
\right)\left[
{\cal T}\left\{H_{I\sigma'}(t')H_{I\sigma'}(t'')\right\}
- \overline{\cal T}\left\{H_{I\sigma'}(t')H_{I\sigma'}(t'')\right\}
\right].
\eea
This is a very happy form. Indeed, when we substitute the expansion
(\ref{eq:Hnor}) in (\ref{eq:Hw3}), we find that contractions of
field operators resulting from normal-ordering the righthand side
will only be of the Feynman type. Moreover, we need only consider
the operator
\bea
\label{eq:Kw}
K_\sigma(t) &=& {\sigma\over 4}\int dt' 
e^{-\sigma|t-t'|}  H_I(t') 
- {i\sigma\over 4}
\int\limits^\infty_\sigma d\sigma' 
{\sigma'^2-\sigma^2\over\sigma'^3}
\int dt' dt'' e^{-\sigma'|t'-t''|} 
\\
\nonumber && \quad \qquad
\times \left(
e^{-\sigma|t-t'|}+e^{-\sigma|t-t''|}
\right)
{\cal T}\left\{H_{I\sigma'}(t')H_{I\sigma'}(t'')\right\},
\eea
since from the Hermiticity of $H_{I\sigma}$ we have
\beq
H_{I\sigma}(t) = K_\sigma(t) + K_\sigma^\dagger(t).
\eeq
The calculation of $K_\sigma$ involves only proper
Feynman propagators, and one need only consider connected
graphs since disconnected graphs will not contribute to
$H_{I\sigma}$ (they are cancelled by identical terms
from $K_\sigma^\dagger$).

We expand $K_\sigma(t)$ as we did for $H_{I\sigma}(t)$
in (\ref{eq:Hnor}) ---
so that $h^{\{p\}}_{I\sigma} = k^{\{p\}}_\sigma +
(k^{\{{\overline p}\}}_\sigma)^\dagger$, where 
$k^{\{{\overline p}\}}_\sigma = k_\sigma 
(p_{{\bar F} + F},...,p_1;-p_{{\bar F}+F+B},...,-p_{{\bar F}+F+1})$
 --- and substitute
in (\ref{eq:Kw}).  The amplitudes $k^{\{p\}}_\sigma$ 
so defined are to be solved
by iteration. One can always organize such an iterative
solution to reproduce the perturbative expansion of Section III,
should one care to. Indeed, by expanding the dressed amplitudes
$h^{\{p\}}_{I\sigma}$ in powers of the undressed
amplitudes $h^{\{p\}}_I$, one recovers the perturbative
scheme of Section III.A.
However, now there are other more interesting
possibilities, for just as with Dyson-Schwinger equations one 
can use a variety of approximations to (\ref{eq:Kw}) in order
to sum up infinite classes of diagrams. We note that (\ref{eq:Kw})
permits a Wick rotation to Euclidean space and thus the normal
power counting applies in determining the counterterm structure.
This does not seem to be true for all choices of the similarity
function $f$.

\subsection{In the Dyson picture}

We can also develop Dyson-Schwinger-like equations for
the intermediate Hamiltonian by introducing a new time scale 
$T>0$ into the integrals in (\ref{eq:S_D}). We define
\beq
S^D(t,T) = {\cal T}\left\{ e^{-i\int_{t-T}^t
dt' H_I(t,t')}\right\},
\eeq
so that
\bea
H^D_I(t,T) &=& -\sigma \sum_{n=0}^\infty i^n \int\limits_{t-T}^t
dt_0 g'(\sigma(t-t_0)) \int\limits_{t-T}^{t_0}
dt_1 g(\sigma(t-t_1))\cdots \\
\nonumber && \qquad \cdots
\int\limits_{t-T}^{t_{n-1}}
dt_n g(\sigma(t-t_n)) [H_I(t_n),\dots [H_I(t_1),H_I(t_0)]\cdots]
\\
\nonumber &=& -\sigma \int\limits_{t-T}^t
dt_0 g'(\sigma(t-t_0)) H_I(t_0)
+ i \int\limits_{t-T}^t dt_1 g(\sigma(t-t_1))
[H_I(t_1),H^D_I(t,t-t_1)],
\eea
which is a linear integral equation.
The usual intermediate Hamiltonian is then obtained from
$H_I^D(t,T)$ by taking the limit $T\to\infty$. 
$H_I^D(t,T)$ depends on $t$ as
an interaction representation operator --- namely,
$H_{I}^D(t,T) = e^{iH_0t} H^D_{I}(0,T) e^{-iH_0t}$ ---
so we expand $H_I^D(t,T)$ in terms of normal-ordered products
of field operators each at time $t$, with coefficents that
depend on the positions of these operators and $T$:
\beq
H_I^D(t,T) = \sum_{\{p\}} h_{ID}^{\{p\}}(T){\cal N}_{\{p\}}(t).
\eeq
${\cal N}_{\{p\}}(t)$ can be expressed as a
normal-ordered product of field operators such as
$\psi_{p_i}(t)\equiv \psi(p_i)e^{-ip_i^0 t}$.
$K_D(t,T)$ is defined analogously
to $K_\sigma(t)$ above, and we find
\bea
\label{eq:Kd}
\sum_{\{p\}} k_D^{\{p\}}(T)
{\cal N}_{\{p\}}(t) &=& \int\limits_0^T dt'
\int d\Gamma G(\Gamma) e^{-\sigma\Gamma t'} \left[ 
{\sigma\Gamma\over 2} \sum_p h_I^{\{p\}} {\cal N}_{\{p\}}(t-t') 
\right. \\
\nonumber && \quad \qquad
-i \left. \sum_{\{p\},\{p'\}} 
h_I^{\{p\}}h_{ID}^{\{p'\}}(t')
{\cal T}\left\{{\cal N}_{\{p\}}(t-t'){\cal N}_{\{p'\}}(t)\right\}
\right].
\eea
The time-ordered product is expanded using Wick's Theorem
in momentum space --- for example,
\beq
{\cal T}\left\{\psi_q(t)\psi^\dagger_{q'}(t')\right\} =
:\!\psi_q(t)\psi^\dagger_{q'}(t')\! : +
 i (2\pi)^4\delta^{(4)}(q-q') S_F(q)\gamma_0 e^{-iq_0(t-t')}.
\eeq
We represent the product of all contractions --- that is, 
the product of $B_q$
boson propagators $D_F(q_i)$, $F_q$ $\psi \psi^\dagger$ propagators
$S_F(q_j)\gamma_0$ and 
${\bar F}_q$ $\psi^\dagger\psi$ propagators
$-S_F(-q_k)\gamma_0$ --- as ${\cal M}_{\{q\}}$, so that we have
\bea
{\cal T}\left\{{\cal N}_{\{p\}}(t-t'){\cal N}_{\{p'\}}(t)\right\}
&=& \delta_{N_q N_{q'}}\sum_{B_q {\bar F}_q F_q} 
\left\{\prod_{i=1}^{N_q} 
(2\pi)^4 \delta^{(4)}(q'_i - {\bar q}_i)\right\} \\
\nonumber && \qquad \times
{\cal M}_{\{q\}} :\!{\cal N}_{\{p/q\}}(t)
{\cal N}_{\{p'/q'\}}(t)\! : e^{-ip^0 t'}. 
\eea
The notation ${\cal N}_{\{p/q\}}$ means the
$N_q = {\bar F}_q + F_q + B_q$ field operators with
$p_i=q_i$ (those corresponding to the contractions)
are dropped from the normal-ordered product 
${\cal N}_{\{p\}}$. 
Expressing the amplitudes as
\beq
k_D^{\{p\}}(T) = \int\limits_0^\infty d\Gamma \,
{1-e^{-(\sigma\Gamma+ip^0)T} \over \sigma\Gamma+ip^0}
 {\widetilde k}_D^{\{p\}}(\Gamma)
\eeq
and similarly for $h^{\{p\}}_{ID}(T)$, after a bit of
manipulation (\ref{eq:Kd}) becomes
\bea
\label{eq:kd1}
{\widetilde k}^{\{p\}}_D(\Gamma) &=&
{\sigma\Gamma\over 2}G(\Gamma)h^{\{p\}}_I  
+ i \int\limits_0^{\Gamma} d\Gamma' G(\Gamma')
\sum_{\{q\}}\sum_{{\bar F}_{p'}F_{p'}B_{p'}} 
{h_I^{\{p',q\}}{\cal M}_{\{q\}}
{\widetilde h}_{ID}^{\{p/p',{\overline q}\}}\!(\Gamma-\Gamma')
\over \sigma(\Gamma-\Gamma')+i(p^0-p^{'0}-q^0)}.
\eea
This equation has a simple graphical interpretation in analogy
with the spurion diagrams of Section III.C.
If we define the dressed amplitudes $k_D^{\{p\}}(\Gamma,\nu)
= 2\pi\delta(p^0 - i\sigma\Gamma - \sigma\nu) 
{\widetilde k}_D^{\{p\}}(\Gamma)$
and $h_{ID}^{\{p\}}(\Gamma,\nu)
= 2\pi\delta(p^0 - i\sigma\Gamma - \sigma\nu)
{\widetilde h}_{ID}^{\{p\}}(\Gamma)$
and the undressed amplitude $h_I^{\{p\}}(\Gamma,\nu)
= 2\pi\delta(p^0 - i\sigma\Gamma - \sigma\nu) G(\Gamma)h_{I}^{\{p\}}$,
then (\ref{eq:kd1}) becomes
\bea
\label{eq:kd2}
k^{\{p\}}_D(\Gamma,\nu) &=& 
{\sigma\Gamma\over 2} h^{\{p\}}_I(\Gamma,\nu)
\\ \nonumber && \quad  
+  \int d\Gamma'\int {d\nu'\over 2\pi}
\sum_{\{q\}}\sum_{{\bar F}'F'B'} 
h_I^{\{p',q\}}(\Gamma',\nu'){\cal M}_{\{q\}}
{h_{ID}^{\{p/p',{\overline q}\}}\!(\Gamma-\Gamma',\nu-\nu')
\over\nu-\nu'-i\varepsilon}.
\eea
If we draw a spurion line as in Figure 4, entering the undressed vertex
with momentum $i\Gamma\lambda^\mu$, carrying momentum
$(-\nu'+i\Gamma-i\Gamma')\lambda^\mu$ from the undressed to the
dressed vertex and exiting the last vertex with momentum
$-\nu\lambda^\mu$, then the total 4-momentum is conserved at
each vertex. 

Figure 4 is a graphical representation of
(\ref{eq:kd2}) for the ${\cal N}^{220}_{\{p\}}$ amplitude. Undressed
vertices are denoted by a dot, and dressed vertices (of
type $k_D$ or $h_D$) are denoted by a bubble. The resemblance
to Dyson-Schwinger equations is evident. The dots indicate that
we have not shown diagrams that modify the quark legs only (counterterm
diagrams from the undressed Hamiltonian and $\sigma$-dependent
mass terms from the dressed Hamiltonian) as well as diagrams that
involve dressed vertices with more than five total quark and
gluon legs. The diagrams for $k_D^\dagger$ are equivalent to
those shown here, except the direction of the
spurion line is reversed. Diagrammatically, it is easy to see
how an iterative solution of these equations, starting with the
undressed vertices as input, will reproduce the peturbative
diagrams of Section III.C. Of course, the diagrams are precisely
equivalent only in the limit $T\to\infty$. Note finally that
because the amplitude $k_D(p_3,-p_2;-p_4,p_1;\Gamma)$ enters
$K_D$ as the coefficient of ${\cal N}^{220}(p_3,-p_2;-p_4,p_1)$,
which is antisymmetric under either $p_1 \leftrightarrow -p_4$ or
$p_3 \leftrightarrow -p_2$ and symmetric under 
$(p_1,-p_4) \leftrightarrow (p_3,-p_2)$ plus complex conjugation,
the diagrams in Fig. 4 represent additional processes --- for
example, the last diagram shown represents processes where the
undressed $qg \to q$ vertex is attached to any of the four
quark legs.

The advantage of having an equation that is linear in the
dressed vertices, as in the present case, is that the undressed
Hamiltonian has a finite number of terms, whereas the dressed
Hamiltonian can have terms with any number of legs. Thus a
diagrammatic expansion for any given dressed vertex based upon
(\ref{eq:Kw}), which is nonlinear in the dressed vertices, will
contain an infinite number of graphs since there can be any number
of internal lines connecting the two dressed vertices. In
contrast, the diagrammatic expansion for any given dressed
vertex based upon (\ref{eq:kd2}) contains a finite number of
graphs, since the number of internal lines is limited by the
undressed vertices.

\subsection{Another example}

As an example of this final formulation of the similarity
transformation, let us set up the calculation of an approximation
to the general quark-antiquark potential in light-front QCD.
We start from the diagrams in Fig. 4.  There will be similar
diagrammatic equations for the dressed
$qg\to q$ and ${\bar q}qg\to {\bar q}q$ that appear in the
equation for the dressed ${\bar q}q \to {\bar q}q$ amplitude.
We may approximate these latter equations as shown in Fig. 5.
According to this approximation, we treat perturbatively 
vertices that change gluon number; in first
approximation the dressed $q\to qg$ is equivalent to the
undressed vertex. Consistent with this approximation, we take
the amplitude ${\bar q}qg\to{\bar q}q$ to be as in Fig. 5. This
corresponds to taking the input potential for the dressed
${\bar q}q\to {\bar q}q$ amplitude to be as in Fig. 6. This
yields the ${\cal O}(g_R^2)$ potential calculated in Section III.D.
(For simplicity, we have neglected to draw the spurion lines
in Figs. 5 and 6.)
The equation for the dressed ${\bar q}q\to{\bar q}q$ vertex
in this approximation is shown in Fig. 7. This looks very much
like the Bethe-Salpeter equation. Here again, we have ignored
corrections to the external legs. Clearly, improving the
approximations for the gluon-number-changing vertices
corresponds to including higher order radiative corrections
to the input potential of Fig. 6.

While we shall not solve for this dressed amplitude here, it
is perhaps instructive to set up the equations in some detail.
From the above, the dressed vertices satisfy
\bea
\label{eq:hT}
h_{ID}^{\{p\}}(t) &=& e^{ip^0t} \lim_{T\to\infty}
\int\limits_0^\infty d\Gamma \int {d\nu\over 2\pi i}
{1-e^{-i\nu T}\over \nu-i\varepsilon} h_{ID}^{\{p\}}(\Gamma,\nu)
\\ \nonumber &=& e^{ip^0t} \int_0^\infty dt'
\int\limits_0^\infty d\Gamma \int {d\nu\over 2\pi}
e^{-i\nu t'} h_{ID}^{\{p\}}(\Gamma,\nu),
\eea
with
\bea
\label{eq:hd2}
h_{ID}^{\{p\}}(\Gamma,\nu) &=& \sigma\Gamma\, h_I^{\{p\}}(\Gamma,\nu)
+ \int\limits_0^\infty d\Gamma' \int {d\nu'\over 2\pi}
\sum_{\{q\}}\sum_{{\bar F}'F'B'} 
 \\ \nonumber && \qquad \times \left\{
{1\over \nu - \nu'- i\varepsilon}
h_I^{\{p',q\}}(\Gamma',\nu') 
{\cal M}_{\{q\}}
h_{ID}^{\{p/p',{\overline q}\}}\!(\Gamma-\Gamma',\nu-\nu')
\right. 
 \\ \nonumber && \qquad\quad  - \left.
{1\over\nu'- i\varepsilon}
h_{ID}^{\{p',q\}}(\Gamma',\nu')
\left[{\cal M}_{\{q\}}\right]^\dagger
h_I^{\{p/p',{\overline q}\}}\!(\Gamma-\Gamma',\nu-\nu')
\right\}.
\eea
To ease our notational burden, let us call the dressed
$q{\bar q}\to q{\bar q}$ vertex $h_{ID}^{220}\equiv \Phi$;
the dressed $q\to qg$ vertex $h_{ID}^{111}\equiv \phi$; and
the dressed $q{\bar q}g\to q{\bar q}$ vertex 
$h_{ID}^{221}\to \Psi$. Undressed vertices will be denoted
by the subscript $0$ --- note that $\Psi_0=0.$ Now the
approximations of Fig. 5 are
\beq
\label{eq:qqg}
\phi(\Gamma,\nu) \sim  \sigma\Gamma \phi_0(\Gamma,\nu);
\eeq
\bea
\label{eq:qqg-qq}
\Psi(\Gamma,\nu) &\sim& i\int\limits_0^\infty d\Gamma' \int {d\nu'\over 2\pi}
\int{d^4q \over (2\pi)^4}
\left\{
{1\over \nu - \nu'- i\varepsilon}
\phi_0(\Gamma',\nu') S_F(q)
\Phi(\Gamma-\Gamma',\nu-\nu') \right.
\\ \nonumber && \qquad + \left.
{1\over\nu'- i\varepsilon}
\Phi(\Gamma',\nu') S_F(q) \phi_0(\Gamma-\Gamma',\nu-\nu') 
\right\}.
\eea
The undressed quark-gluon vertex is\cite{ZH}
\bea
\phi_0(\Gamma,\nu) &=& G(\Gamma){\bar\phi}_0 (2\pi)^4\delta^{(4)}
(p_3^0-p_1^0-q_0-(i\Gamma+\nu)\lambda)
\\ \nonumber
{\bar\phi}_0(p_3,p_1,q) &=& -g_R T^a \left\{ 2{q_\perp^i\over [q^+]}
- {\sigma_\perp\cdot p_{3\perp}-im_F\over [p_3^+]}\sigma^i_\perp
- \sigma^i_\perp
{\sigma_\perp\cdot p_{1\perp}+im_F\over [p_1^+]}
\right\}.
\eea
The undressed $q{\bar q}\to q{\bar q}$ vertex is
\bea
&\Phi_0(\Gamma,\nu) = \int d\Gamma_0 G(\Gamma_0)
d\Gamma_1 G(\Gamma_1) \delta(\Gamma-\Gamma_0-\Gamma_1)
{\bar\Phi}_0 (2\pi)^4\delta^{(4)}(p^0-(i\Gamma+\nu)\lambda)&
\\ \nonumber
&{\bar\Phi}_0(p_3,-p_2;-p_4,p_1) = 4g^2_R
 T^a T^a {1\over [p^+_3-p^+_1]^2}.&
\eea
We have seen these undressed vertices in the perturbative calculation
of Sec. III.D. 

The next to the last graph in Fig. 4 is
\beq
\label{eq:f4a}
\int\! d\Gamma' \int\! {d\nu'\over 2\pi} \int\!{d^4q\over (2\pi)^4}
{1\over\nu-\nu'-i\varepsilon}
\phi_0^\dagger(\Gamma',\nu') iD(q)\phi(\Gamma-\Gamma',\nu-\nu').
\eeq
Using the approximation (\ref{eq:qqg}),
this becomes
\bea
\label{eq:f4ai}
\Phi^a_0(\Gamma,\nu) 
&\equiv& -\int\! d\Gamma_0 G(\Gamma_0)d\Gamma_1 G(\Gamma_1)
\delta(\Gamma-\Gamma_0-\Gamma_1)
\int\! {d\nu'\over 2\pi i} 
 {\sigma\Gamma_0 \over\nu-\nu'-i\varepsilon} \\
\nonumber && \qquad \times
{\bar\phi}_0^\dagger(-p_4,-p_2,-q) D(-q) {\bar\phi}_0(p_3,p_1,q)
(2\pi)^4 \delta^{(4)}(p - (i\Gamma+\nu)\lambda), 
\eea
where $q = p_2-p_4+(i\Gamma_1+\nu')\lambda$. There is a similar contribution
to $\Phi$ from the term $k_{ID}^\dagger$. Indeed, it is identical
after the replacement $q\to \tilde{q}= 
p_3 - p_1 - (i\Gamma_1+\nu')\lambda$ ---
this corresponds to reversing the direction of the spurion line.
We write this as $\Phi^b_0 = \Phi^a_0|_{q\to\tilde{q}}$. 
After integration over $\nu'$, 
the sum of these terms and the undressed vertex $\Phi_0$ gives
the input potential of Fig. 6, namely:
\bea
V(\Gamma,\nu) &=& \sigma\Gamma\Phi_0(\Gamma,\nu) + \Phi_0^a(\Gamma,\nu)
+ \Phi_0^b(\Gamma,\nu) \\
\nonumber &=& \sigma\int d\Gamma_0 G(\Gamma_0)\int d\Gamma_1 G(\Gamma_1)
\Gamma_0 \delta(\Gamma-\Gamma_0-\Gamma_1) \\
\nonumber && \qquad \times {\overline V}_{\{p\}}(i\Gamma_1+\nu)
(2\pi)^4 \delta^{(4)}(p^0 - (i\Gamma + \nu)\lambda),
\eea
where
\bea
{\overline V}_{\{p\}}(\nu+i\Gamma_1) &=& 
{\bar\Phi}_0(q) + {\bar\Phi}_0({\tilde q}) -
{\bar\phi}_0^\dagger(-p_4,-p_2,-q) D(q) {\bar\phi}_0(p_3,p_1,q)
\\ \nonumber && \qquad\qquad
- {\bar\phi}_0^\dagger(-p_4,-p_2,-{\tilde q}) D({\tilde q}) 
{\bar\phi}_0(p_3,p_1,{\tilde q}),
\eea
with $q = p_2-p_4+(\nu+i\Gamma_1)\lambda$ and
${\tilde q} = p_3-p_1-(\nu+i\Gamma_1)\lambda$.
Substituting this into (\ref{eq:hT}), we recover the order
$g_R^2$ potential of Sec. III.D, as indeed we must. Thus the
first and fourth terms on the righthand side of the equation in
Fig. 4 (plus their conjugate graphs)
sum to give the first term on the righthand side in
Fig. 7.

The last graph in Fig. 4 is
\beq
\label{eq:f4}
\int\! d\Gamma' \int\! {d\nu'\over 2\pi} \int\!{d^4q\over (2\pi)^4}
\int\!{d^4q'\over (2\pi)^4} {1\over\nu-\nu'-i\varepsilon}
\phi_0(\Gamma',\nu')D(q)S_F(-q')\Psi(\Gamma-\Gamma',\nu-\nu').
\eeq
Using the approximation (\ref{eq:qqg-qq}), the integrand in
(\ref{eq:f4}) becomes
\bea
\label{eq:f4i}
&& -\int\! d\Gamma'' \int\! {d\nu''\over 2\pi i} \int\!{d^4q''\over (2\pi)^4}
 {1\over\nu-\nu'-\nu''-i\varepsilon} \\
\nonumber && \qquad \times
\left[ \phi_0^{-q',-p_2,-q}(\Gamma',\nu')\right]^\dagger D(q)S_F(-q')
\phi_0^{q'',p_1,q}(\Gamma'',\nu'') S_F(q'')
\Phi(\Gamma-\Gamma'-\Gamma'',\nu-\nu'-\nu'')
\eea
plus a similar term in which the gluon couples leg $-p_2$ to $p_3$.
Other terms are either ignored because they only correct the external
legs or else not distinct because of the symmetries of the amplitude.
The term given in (\ref{eq:f4i}) may be summed with the second graph
on the righthand side of the equation in Fig. 4 to give 
\beq
\label{eq:hDb}
\int\! d\Gamma' \int\! {d\nu'\over 2\pi} \int\!{d^4q\over (2\pi)^4}
\int\!{d^4q'\over (2\pi)^4} 
V_{q,-p_2;-q',p_1}(\Gamma',\nu')
{S_F(q)S_F(-q')\over\nu-\nu'-i\varepsilon}
\Phi_{p_3,-q';-p_4,q}(\Gamma-\Gamma',\nu-\nu')
\eeq
This is the second graph on the righthand side in Fig. 7. The
remaining graphs in Fig. 7 result from (\ref{eq:hDb}) by
complex conjugation (reversing the spurion line) and by
the substitution $-p_2 \leftrightarrow -p_4$.
We have, finally,
\bea
\label{eq:Phi}
\Phi_{p_3,-p_2;-p_4,p_1}(\Gamma,\nu) &=& 
V_{p_3,-p_2;-p_4,p_1}(\Gamma,\nu) +
\int\! d\Gamma' \int\! {d\nu'\over 2\pi} \int\!{d^4q\over (2\pi)^4}
\int\!{d^4q'\over (2\pi)^4} \\
\nonumber &&  \times \left\{
V_{q,-p_2;-q',p_1}(\Gamma',\nu')
{S_F(q)S_F(-q')\over\nu-\nu'-i\varepsilon}
\Phi_{p_3,-q';-p_4,q}(\Gamma-\Gamma',\nu-\nu') \right.\\
\nonumber && \quad +
\Phi_{q,-p_2;-q',p_1}(\Gamma',\nu')
{S_F(q)S_F(-q')\over\nu'-i\varepsilon}
V_{p_3,-q';-p_4,q}(\Gamma-\Gamma',\nu-\nu')\\
\nonumber && \quad -
V_{q,q';p_4,p_1}(\Gamma',\nu')
{S_F(q)S_F(q')\over\nu-\nu'-i\varepsilon}
\Phi_{p_3,p_2;q',q}(\Gamma-\Gamma',\nu-\nu')\\
\nonumber && \quad - \left.
\Phi_{q,q';p_4,p_1}(\Gamma',\nu')
{S_F(q)S_F(q')\over\nu'-i\varepsilon}
V_{p_3,p_2;q',q}(\Gamma-\Gamma',\nu-\nu')\right\}.
\eea
It would be interesting to approximate $V$ further by just taking
the long range (small ${\bar q}$) part calculated in Sec. III.D and
substituting this in (\ref{eq:Phi}). The divergence in $V$ will be
cancelled by divergences in the self-energy corrections to the external
legs (see Sec. III.D) --- corrections which we have ignored here but
which must be included if one wants to solve this Bethe-Salpeter-like
equation in a consistent manner. This paper is long enough already,
and therefore such a calculation will not be undertaken here.

\section{Closing Arguments}

In this paper we have developed a similarity renormalization
framework for the study of strongly interacting field theories. 
The basic view of this approach is that Hamiltonian methods
will prove most fruitful for the solution of such theories.
The generic theory for which our framework has been developed
has interactions that are weak asymptotically but grow large
enough at low energy to invalidate perturbation theory. 
QCD is of course the theory that we particularly have in mind. 
The basic idea behind the similarity renormalization scheme is
that the Hamiltonian of such a theory can be transformed into
that of an effective many-body theory where states that differ
greatly in free energy are essentially uncoupled. This resulting
many-body Hamiltonian can then be solved with recourse to the
many available approximate methods such as variational approaches,
trial wave functions, iterative techniques and numerical basis
function approaches. The similarity transformation achieves this
goal by smoothly eliminating interactions between states whose
free energies differ by an amount greater than an arbitrary
scale $\sigma$ introduced by the transformation. The result
of this elimination is to create new interactions between states
that differ in energy by an amount less than $\sigma$. The
assumption is that from these new interactions one can
extract potentials that describe the relevant physics.

Clearly, no physical result can depend on the renormalization
scale $\sigma$. However, the calculations can only be approximately
carried out for most systems of interest, and so one will need to
choose a value of
$\sigma$ that makes the approximations as valid as possible.
Computations in this framework involve two steps: first, the 
similarity transformed Hamiltonian is calculated; second, this
Hamiltonian is solved. The first step is most easily done 
perturbatively, which requires that $\sigma$ be large since only
at high energy is the coupling small (by assumption). The second
step is most easily done when as few states as possible are
coupled strongly, which requires that $\sigma$ be as small as
possible. Inevitably, one must make a compromise and choose
$\sigma$ in some intermediate range. If the Dyson-Schwinger-like
equations developed in the last section can be implemented, it
may be possible to go beyond the perturbative calculation of
the dressed Hamiltonian so that $\sigma$ can be lowered to
values for which perturbation theory begins to break down. 
However, in most
cases it will probably be better to keep $\sigma$ large enough
that the Hamiltonian can be dressed peturbatively, since the tools
available for the second step are generally more powerful.

That the elimination of couplings between states separated by large
energy differences turns a field theory into a many-body theory
is clear if the free particles are massive. If $\sigma$ is lowered
to just above the particle masses, then creating or destroying
a particle will cause an energy difference of order $\sigma$. This
means interactions that change particle number will be suppressed.
If the particles are light or massless, however, as in QCD, it does
not follow that the similarity transformation described here will
suppress interactions that change particle number. Indeed, one
might consider it better to have the similarity function depend
on differences in particle number rather than on differences
in energy. However,
such a transformation is unlikely to be feasible in perturbation
theory, for one no longer has a means of eliminating only 
high energy degrees of freedom. 
Moreover, one would lose the advantage that small
energy denominators are avoided. 
Instead, one can use the similarity transformation in a
light-front formulation of a theory such as QCD. We shall argue
in the following that this change of formulation of the theory
will provide a means of turning even QCD into a many-body problem.

The use of light-front coordinates in field theory is not new (see
Ref. [3] for a review and an exhausting --- if 
perhaps not quite exhaustive --- list
of references). Such calculations are considered difficult, partly
because they are unfamiliar, but also because they really are difficult.
One of the major difficulties with the use of light-front coordinates
is that Lorentz covariance is no longer manifest once one chooses a
particular direction in defining the light-front time. Another problem
is that calculations are only feasible for a particular gauge choice,
the light-cone gauge. These problems severely complicate the
renormalization of divergences, and this has slowed advance in
light-front computations. The fact that in our similarity renormalization
scheme the calculation of the dressed Hamiltonian can be expressed in
terms of Feynman diagrams is therefore significant. It means that
divergences can be handled with dimensional regularization, the
infrared singularities can be handled with the Mandelstam-Leibbrandt
prescription, and the
counterterm structure is then that of covariant formulations. In particular,
renormalization of divergences for QCD in equal-time coordinates 
and the light-cone gauge is well
understood (see, for example, Refs. [10-11]), and we can rely on such
prior work in determining the dressed Hamiltonian in the intermediate
representation. There are of course other, more technical difficulties
resulting from the choice of light-front coordinates (for example,
rotations about the transverse axes are no longer kinematical); but
these should not be prohibitive. Indeed, there are also
some technical advantages to the use of light-front as compared to
equal-time coordinates (for example, now boosts are kinematical).

The essential reasons for using light-front coordinates follow from
the free particle dispersion relation
\beq
\label{eq:dr}
p^- = \frac{p_\perp^2 + m^2}{p^+}.
\eeq
This implies that all on-shell particles (and antiparticles) have
have longitudinal momentum $p^+\ge 0$ and all physical states have
total longitudinal momentum $P^+\ge 0$. The vacuum has $P^+=0$ and
is therefore built only from particles with $p^+=0$. From (\ref{eq:dr})
we see that such particles have infinite (light-front) energy unless
$p_\perp^2=0$ and $m^2=0$. The similarity transformation will decouple
the infinite energy states from the low-energy physics, and the
computation of the vacuum state is then greatly simplified. Indeed, if
we give gluons a small mass, all particles with $p^+=0$ have infinite
energy, and the vacuum state of the dressed Hamiltonian
is just the trivial free vacuum state. Likewise, in building any
bound state of the dressed Hamiltonian, we can rely on the fact
that the total longitudinal momentum $P^+$ must be built up from
constituent longitudinal momenta $p^+_i\ge 0$. If the energy of
each constituent is to be less than $\sigma$, we must have
$p^+_i \ge (p_\perp^2+m^2)/\sigma$ for all $p^+_i$, and thus in
effect we have a bound on the number of particles that can make
up a state. Again, the only particles that escape this lower bound
on the longitudinal momentum are (massless) gluons with $p_\perp^2\sim 0$,
but one can hope that such highly transversely localized particles
will have little contribution to low-energy states. In any case,
it is clear that effects associated with the quark-gluon sea in the
equal-time formulation of QCD will be largely, if not totally, 
replaced by interactions in the dressed Hamiltonian after a
similarity transformation in the light-front formulation.

But the most interesting result of the light-front formulation
is that longitudinal and transverse variables scale differently
(this has been repeatedly emphasized by Wilson\cite{W}). This can
be seen from the dispersion relation (\ref{eq:dr}) and has been
discussed in Sec. III.D. Thus the similarity renormalization of
light-front Hamiltonians depends on the energy scale $\sigma$
through both a mass scale $\mu$ and a longitudinal momentum
scale $\rho$. The mass scale $\mu$ may be taken to be that introduced
by dimensional regularization. The QCD coupling $g_R$ runs with
this scale, and $\mu$ must be taken large to keep $g_R$ small
enough for perturbation theory to remain valid. However, the
similarity function $f$ runs with the ratio $\sigma=\mu^2/\rho$;
and thus, even though the mass scale $\mu$ be large,
we can suppress couplings between states with 
energy differences above a given value by making the longitudinal
scale $\rho$ large enough. We have seen that light-front
energy increases with particle number;
and therefore at large $\rho$
the similarity transformation will suppress number changing
interactions in the dressed Hamiltonian, thus turning the
solution of QCD bound states into a many-body problem. 
That the mass scale $\mu$
may be nevertheless kept large provides the possibility that the
transformation may be accurately calculated in perturbation
theory. That Dyson's intermediate representation is a particular
similarity renormalization scheme that may be expressed in 
terms of ordinary Feynman graphs means that perturbative light-front
QCD computations are feasible. These computations
should yield a dressed Hamiltonian that is considerably closer to 
strong-interaction phenomenology than the usual free QCD Hamiltonian.
That we find a confining potential already at tree level may be taken as a
positive indication of the utility of this approach.

\centerline{\bf Acknowledgements}

I have benefited from past
discussions with Ken Wilson, Avaroth Harindrinath and Robert Perry. 
I thank J.S. Walhout for first pointing out Ref.[6] to me.





\textheight30cm

\begin{figure}[htp]
\caption{\protect\label{fig1}
Doubled Feynman graph}
\vskip16.2cm
\hskip-3.3cm
\includegraphics{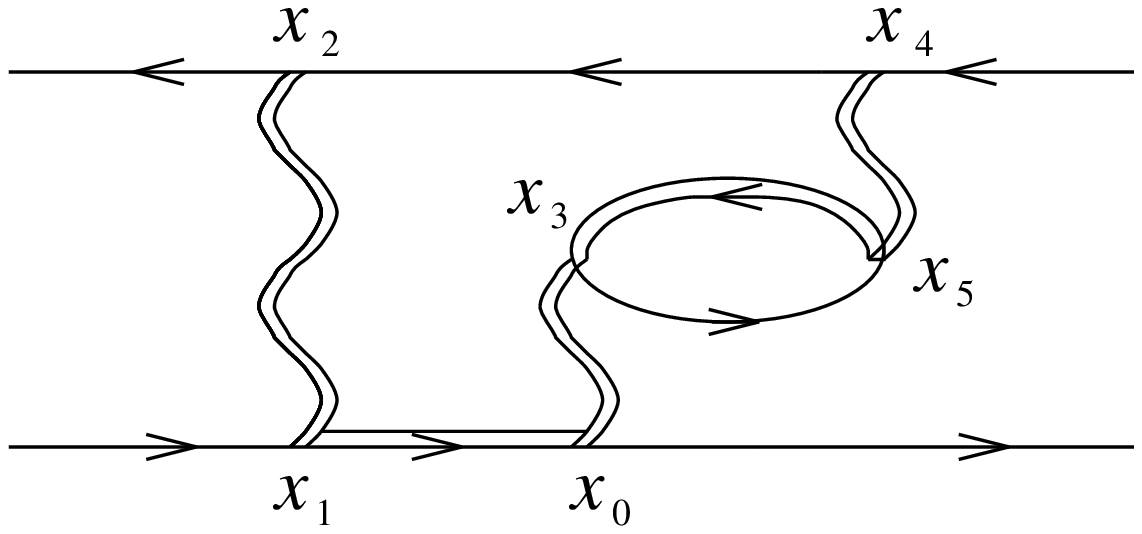}
\vskip-9.0cm
\end{figure}

\vskip-2cm

\begin{figure}[htp]
\caption{\protect\label{fig2}
Spurion graph corresponding to Figure \ref{fig1}}
\vskip16.8cm
\hskip-3.3cm
\includegraphics{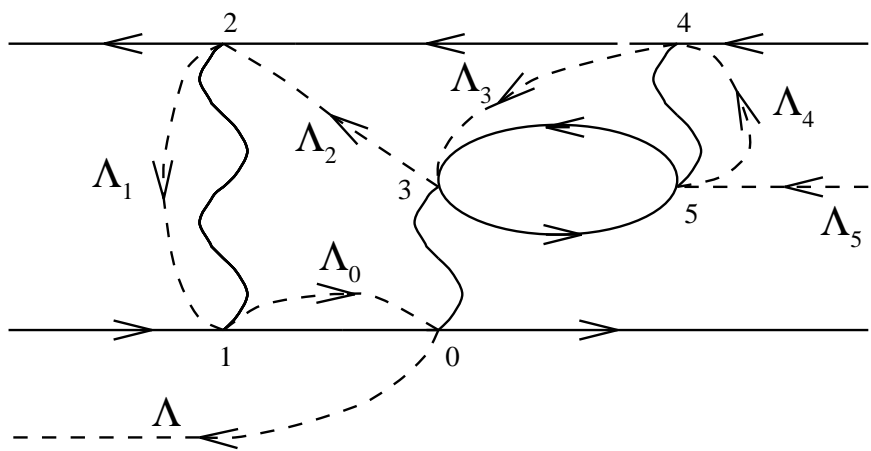}
\vskip-10.5cm
\end{figure}

\newpage



\,

\begin{figure}[htp]
\caption{\protect
Order $g^2_R$ 
diagrams contributing to
$q\bar q\to q\bar q$ }
\vskip22.5cm\hskip-3.4cm 
\includegraphics{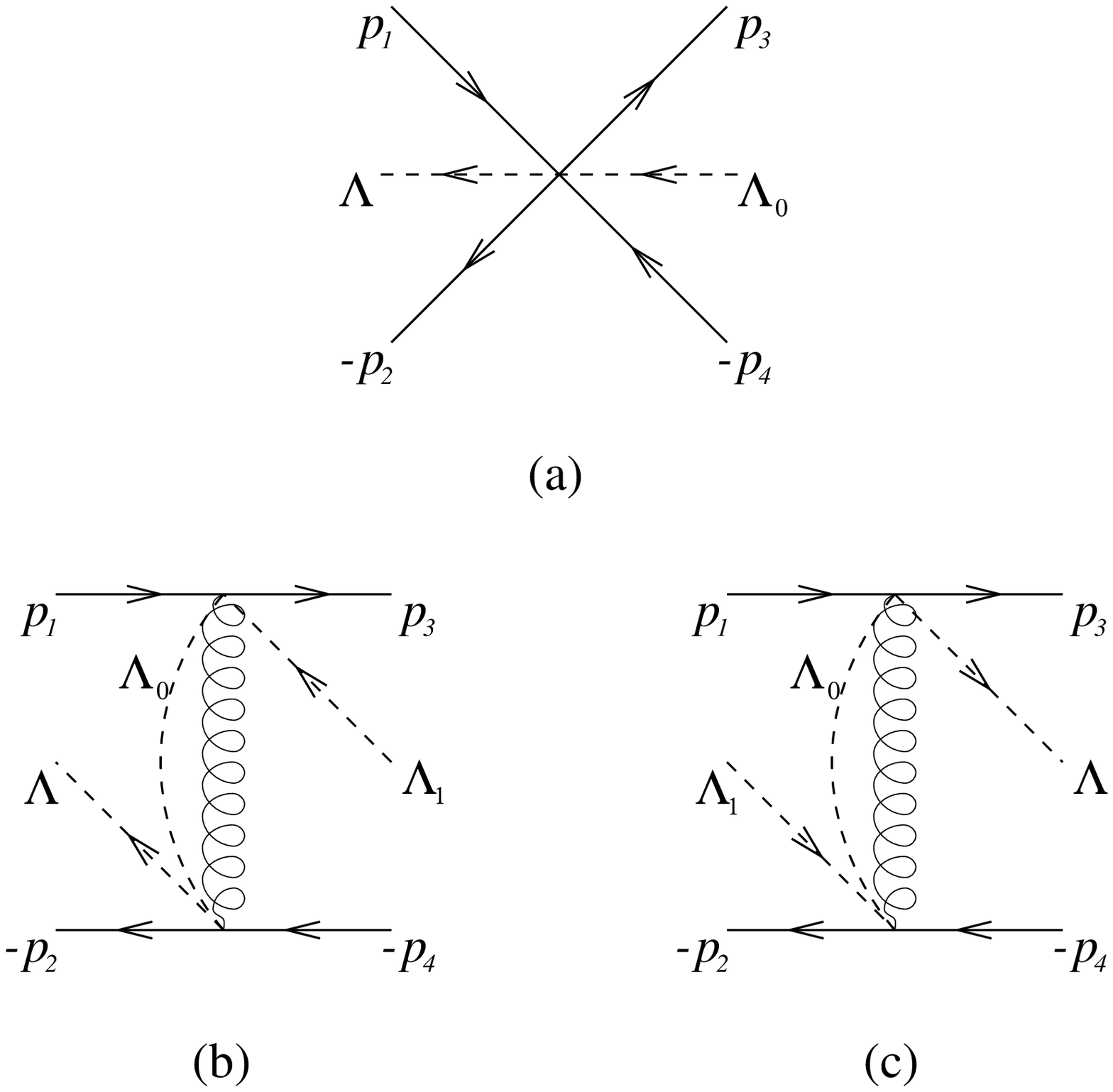} 
\vskip-3.0cm
\end{figure}

\newpage

\begin{figure}[htp]
\caption{\protect
Diagrammatic equation for general
$q{\bar q}\to q{\bar q}$ vertex}
\vskip24.5cm\hskip-3.1cm 
\includegraphics{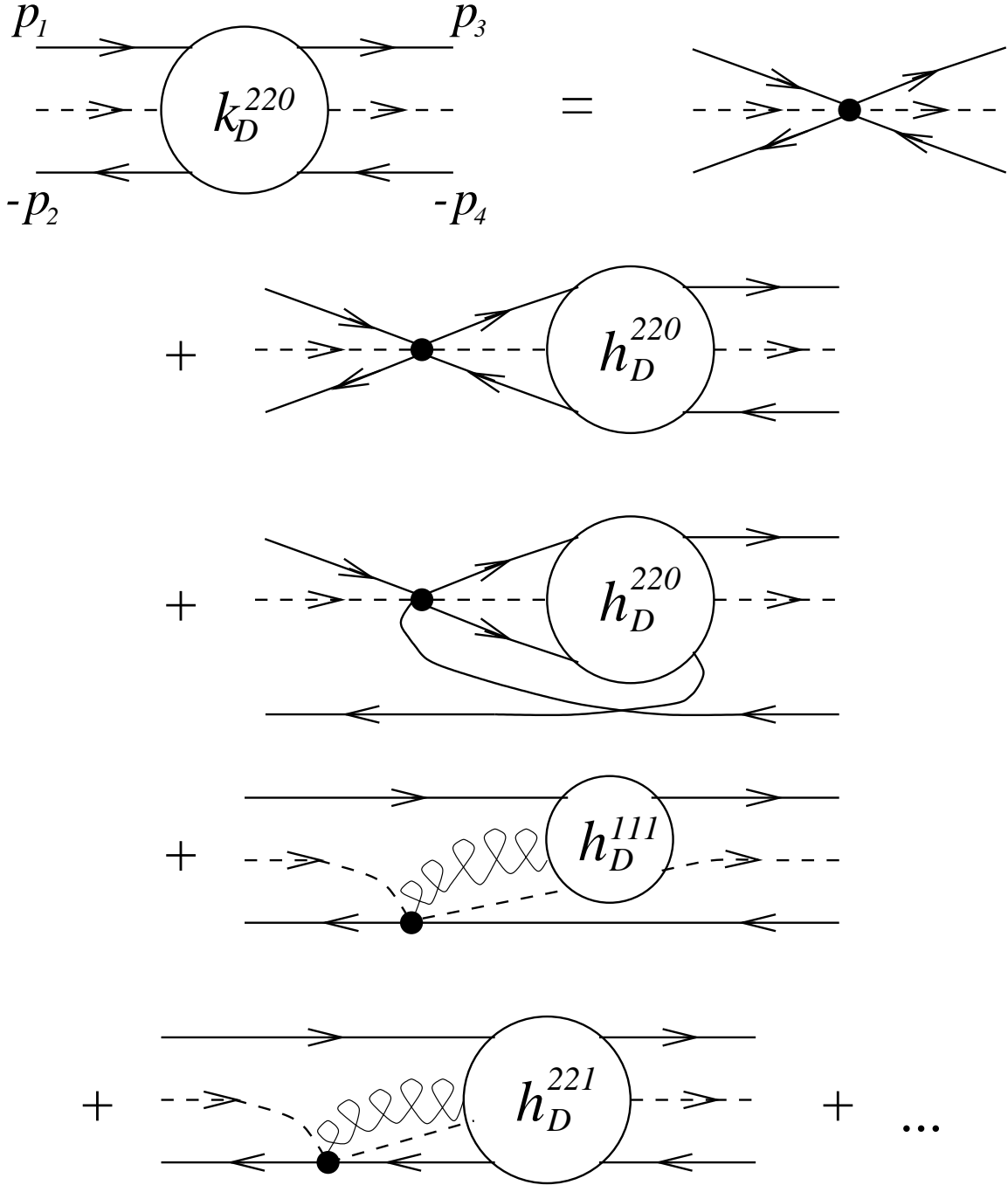} 
\vskip-3.0cm
\end{figure}

\newpage

\begin{figure}[htp]
\caption{\protect
Approximations for 
$q\to qg$ and $q{\bar q}g\to q{\bar q}$
vertices}
\vskip19.0cm\hskip-2.8cm 
\includegraphics{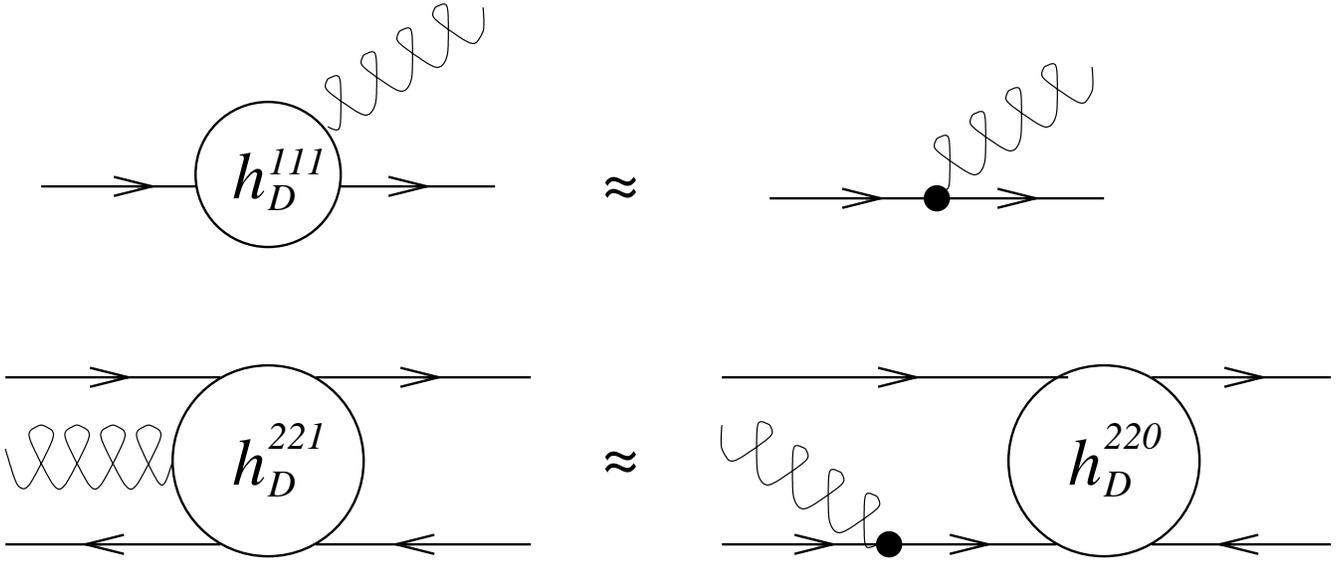} 
\vskip-10.9cm
\end{figure}

\begin{figure}[htp]
\caption{\protect
Definition of input potential for approximation
to $q{\bar q}\to q{\bar q}$ vertex}
\vskip16.3cm\hskip-3.1cm 
\includegraphics{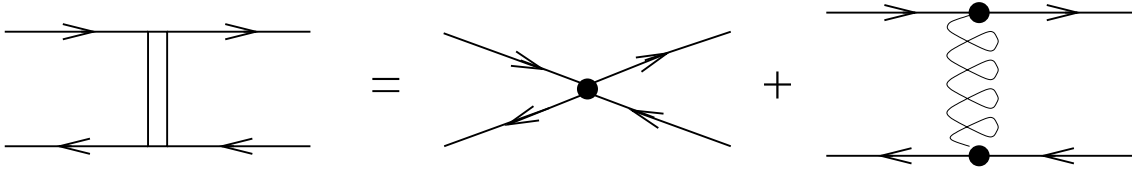} 
\vskip-3.0cm
\end{figure}

\newpage

\begin{figure}[htp]
\caption{\protect
Bethe-Salpeter-like approximation to
$q{\bar q}\to q{\bar q}$ vertex}
\vskip24.5cm\hskip-3.2cm 
\includegraphics{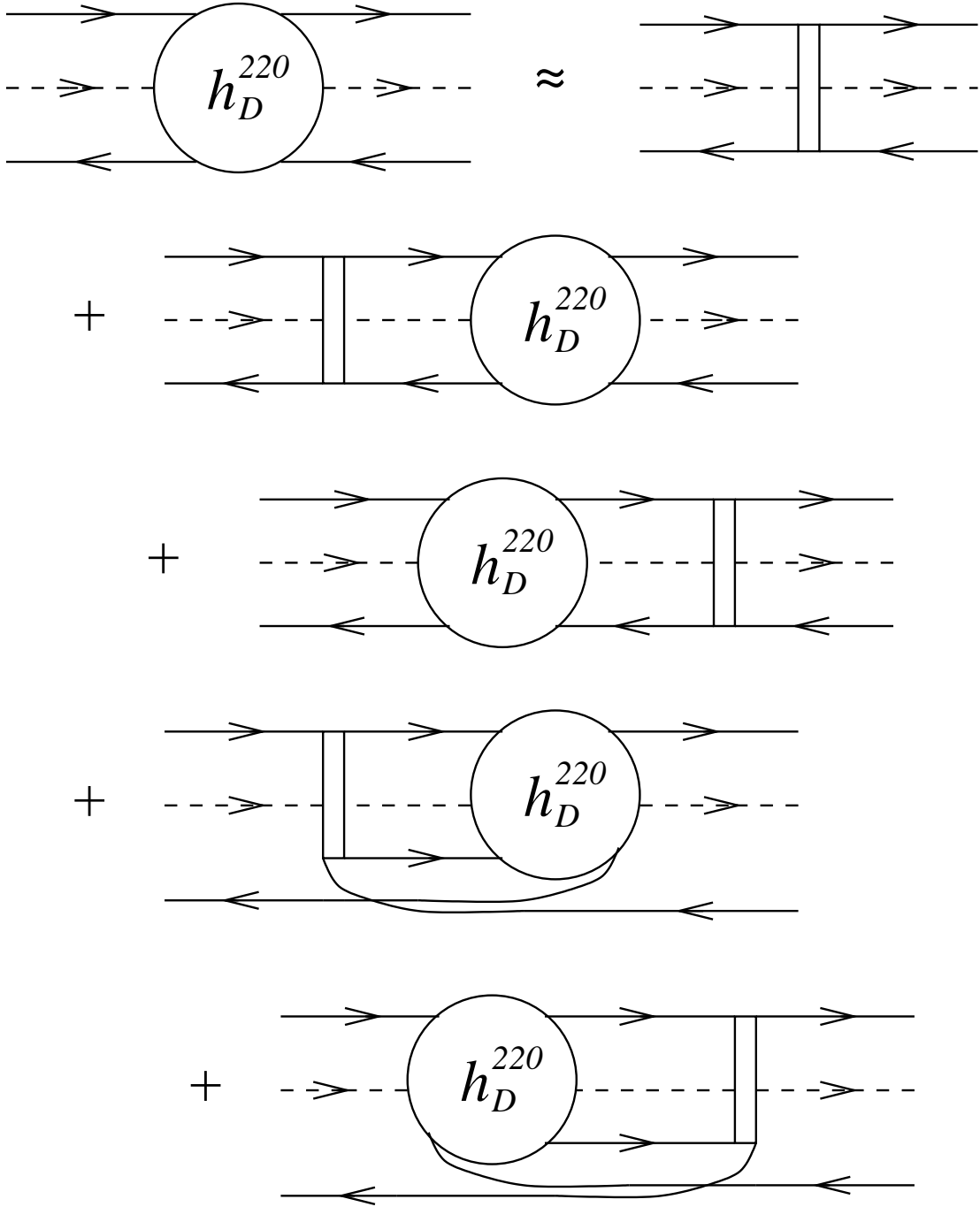} 
\vskip-3.0cm
\end{figure}

\end{document}